\documentclass[aps,prx,twocolumn,nofootinbib,tightenlines,superscriptaddress,nopacs,amsmath,amssymb,final,letterpaper]{revtex4-1}

\usepackage[utf8]{inputenc}
\usepackage{calc}
\usepackage{graphicx}
\usepackage{amsmath,amssymb,amsthm}
\usepackage{bm}
\usepackage{color}
\usepackage[dvipsnames]{xcolor}
\usepackage{enumitem}
\usepackage{float}
\usepackage[ruled,linesnumbered,noline]{algorithm2e}
\SetNlSty{textbf}{}{\hspace{0.5em}}

\DeclareMathOperator*{\argmax}{arg\,max}

\newcommand{\abs}[1]{\vert #1\vert}
\usepackage[normalem]{ulem}

\usepackage[colorlinks=true,linkcolor=blue,urlcolor=blue,citecolor=blue,anchorcolor=blue]{hyperref}

\makeatletter
\g@addto@macro{\UrlBreaks}{\do\-\do\_}
\makeatother
\begin{document}
\title{Scalable detection of higher-order interactions in network data}

\author{Yingbang Zang}
\thanks{These authors contributed equally to this work}
\affiliation{Institute of Data Science, University of Hong Kong, Hong Kong SAR, China}

\author{Yanting Zhang}
\thanks{These authors contributed equally to this work}
\affiliation{Institute of Data Science, University of Hong Kong, Hong Kong SAR, China}

\author{Alec Kirkley}
\email{alec.w.kirkley@gmail.com}
\affiliation{School of Computing and Data Science, University of Hong Kong, Hong Kong SAR, China}

\begin{abstract}
Complex systems are routinely measured and represented through pairwise networks, even when the underlying interactions involve more than two units at once. Recovering this latent hypergraph structure from pairwise measurements is a fundamental inverse problem, but as the space of candidate hyperedges grows exponentially with system size, scalable hypergraph reconstruction at arbitrary interaction orders is out of reach for existing methods. Here we cast hypergraph reconstruction as a local signal-to-noise discrimination problem and use this locality to build a fast algorithm that reconstructs hypergraphs up to any interaction order. Across diverse synthetic and real-world systems our method achieves a high recovery accuracy of latent hypergraph structure while reconstructing hypergraphs up to orders of magnitude more quickly than current approaches. Our approach also yields an information-theoretic detectability boundary that sharply predicts which latent hyperedges are recoverable from pairwise measurements.
\end{abstract}

\maketitle

\section{Introduction}
\label{sec:intro}
Complex systems, in which large numbers of interacting units give rise to emergent collective behavior, are often represented as networks, with edges connecting pairs of units~\cite{newman2003structure}. Although graphs can encode arbitrary higher-order dependencies in system dynamics~\cite{peixoto2026graphs}, a broad range of systems across ecology~\cite{werner1992individual,grilli2017higher}, neuroscience~\cite{yu2011higher,giusti2016two}, and social systems~\cite{benson2018simplicial} have been described in terms of interactions or statistical dependencies among groups of more than two nodes. Such systems are often represented as hypergraphs, in which nodes are joined in groups of arbitrary size~\cite{battiston2020networks,battiston2021physics,bianconi2021higher,bick2023higher}, and this higher-order structure has been proposed as a structural explanation for collective phenomena such as abrupt contagion and explosive synchronization~\cite{iacopini2019simplicial,millan2020explosive,gambuzza2021stability}. Realizing the potential of hypergraphs for complex systems analyses requires knowledge of the underlying higher-order interaction structure, which is rarely observed directly, and so this structure must be inferred from observable data~\cite{battiston2026physics,young2021reconstruct}.

When only pairwise edges are considered, network reconstruction is a relatively mature field, with methods based on auxiliary networks~\cite{yu2006reconstruct,wu2008identification}, sparse regression ~\cite{wang2016reconstruct,nitzan2017reconstruct,jose2017inference,mei2018reconstruct,topal2023reconstruct}, statistical inference~\cite{newman2018reconstruct,tiago2018reconstruct,tiago2019reconstruct,ma2020reconstruction,tiago2025reconstruct,qin2025reconstructability}, and machine learning~\cite{amitava2021inference,zhang2022GNN}. By comparison, higher-order reconstruction remains at an early stage, with most existing approaches inferring higher-order structure from node-level time-series data~\cite{wang2022reconstruct,fan2022reconstruct,malizia2024reconstruct,zang2024reconstruct,rahimi2024reconstruct,delabays2025reconstruct,yang2026reconstruct}. The main obstacle to applying higher-order reconstruction methods in practice is the combinatorial growth of the latent interaction space, since with unrestricted interaction orders the number of possible unobserved hyperedges scales exponentially with the number of nodes. This exponential growth makes exhaustive enumeration computationally intractable, even for modestly sized networks. In the time series setting, several approaches mitigate this combinatorial burden by restricting the candidate space, for example through simplicial closure~\cite{wang2022reconstruct,yang2026reconstruct} or through dependencies between lower- and higher-order structure~\cite{zang2024reconstruct}. Such strategies reduce computation considerably, but this tractability is obtained by restricting the interaction orders in advance or by conditioning on specific structural relationships across orders. 

There are also a few existing methods dealing with directly measured pairwise network observations~\cite{young2021reconstruct,lizotte2023reconstruct,wegner2014subgraph,wegner2024nonparametric,wegner2024motifs,wang2024graphs}. The Bayesian method of Ref.~\cite{young2021reconstruct} assumes an exactly observed network in which every latent interaction appears as a fully connected clique, so that candidate hyperedges are the cliques of the observed graph. Inference in this case is a joint problem over combinations of these hyperedge candidates, explored by Metropolis--Hastings sampling in which each acceptance depends on the full current configuration. Although the method is theoretically elegant, in practice its memory and runtime requirements grow rapidly with the number of maximal cliques in the observed graph, so that the method does not scale easily to beyond a few thousand nodes. Ref.~\cite{lizotte2023reconstruct} relaxes the exact-observation assumption by modeling uncertain pairwise measurements, but restricts candidate groups to two or three nodes. The motif cover method of Ref.~\cite{wegner2024nonparametric} instead allows interactions to take the form of arbitrary connected motifs and casts reconstruction as maximum a posteriori inference over subgraph covers, solved approximately by a greedy heuristic. Because the number of motif types grows super-exponentially with size and subgraph discovery becomes expensive, inference is restricted to a finite set of small motifs. To our knowledge, there is no existing hypergraph reconstruction method that requires no maximum interaction order and no fixed set of candidate structures for inference. 

The minimum description length (MDL) principle~\cite{rissanen1978modeling,grunwald2007minimum}, which advocates for the model providing the shortest description of the data in terms of bits of information, provides a principled regularization criterion for including arbitrarily large hyperedges in higher-order network reconstruction. MDL has proven effective in a range of inference problems on pairwise graphs, including network clustering~\cite{peixoto2017nonparametric,peixoto2019bayesian,morel2024bayesian}, network ensemble comparison~\cite{hebert2024network}, graph similarity~\cite{coupette2021graph}, core-periphery and hub detection~\cite{gallagher2021core,kirkley2024hubs} and analysis of network populations~\cite{kirkley2023compressing}. It has also provided the basis for a number of inferential techniques that identify the most critical interaction orders or hyperedges present in a hypergraph whose hyperedges have been measured directly~\cite{kirkley2025structural,kirkley2026hypergraphbackboning}, and for inferring hypergraph representations of temporal interaction data~\cite{kirkley2024dynamic}. 

Here we cast the reconstruction of latent hyperedges from pairwise observations as an MDL problem in which we compare the shortest description of the observed network with no higher-order structure against its description under a candidate hypergraph. The difference between these two descriptions decomposes into a cost (the bits needed to single out the hyperedge among the combinatorially many alternatives) and a gain (the bits saved by encoding its internal edge pairs against the background density) which takes the form of a Kullback--Leibler (KL) divergence and decouples across hyperedges in the latent higher-order network. A candidate hyperedge is retained only when its gain exceeds its cost, which occurs only when the network is better compressed when including the latent hyperedge. Because of the locality of this MDL criterion, the interaction space can be explored adaptively through local search rather than exhaustive enumeration, with latent hyperedge candidates growing and shrinking during the search process to allow their memberships to be inferred directly from the data rather than fixed beforehand. This formulation enables scalable reconstruction to arbitrarily large interaction orders. Our objective further yields a detectability boundary characterizing when a hyperedge is supported by the observed pairwise edge structure, and we show that the MDL specification cost of a hyperedge falls at the information theoretic scale of the planted dense subgraph problem, so that, for an interaction of growing size, this boundary is statistically optimal at leading order. We test the framework on synthetic and empirical networks, finding accurate reconstruction across a broad range of higher-order structures and validating the predicted detectability transition. In practice, the method runs up to orders of magnitude faster than existing approaches on a wide range of example networks.

\begin{figure*}[t]
    \centering
    \includegraphics[width=\textwidth]{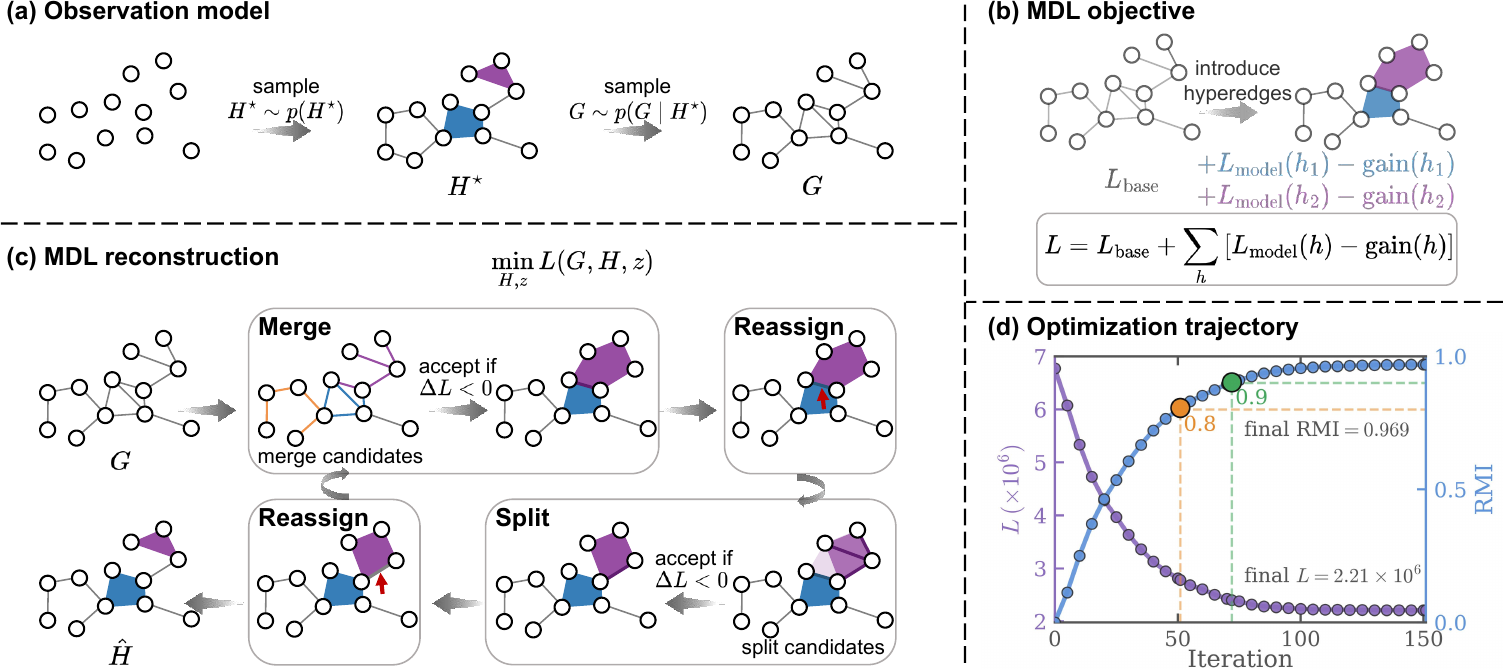}
    \caption{\textbf{Reconstructing latent hyperedges from pairwise observations.}   
    (a) Observation model. The underlying system gives rise to a latent higher-order network $H^\star$, of which only the pairwise projection $G=(V,E)$ is observed.
    (b) MDL objective. Retained hyperedges require structural specification costs but provide compression gains relative to the pairwise noise background, giving the gain--cost objective of Eq.~\eqref{eq:additive_total_main}.
    (c) MDL reconstruction. Seed-centered merge--reassign--split--reassign updates modify the structure of the latent hypergraph, and an edit is accepted when it decreases the description length.
    (d) Optimization trajectory for the coauthorship dataset DBLP 1990 ($N=57{,}247$, $|E|=518{,}454$, with runtime $54.9\,\mathrm{min}$), showing the description length $L$ (purple, left axis) and the reconstruction accuracy measured by the reduced mutual information (RMI)~\cite{jerdee2025RMI} (blue, right axis) versus search iteration. Orange and green markers indicate the iterations at which RMI first reaches 0.8 and 0.9, respectively. RMI is shown only as a diagnostic and is not available to the algorithm, which monitors the description length.}
    \label{fig:fig1}
\end{figure*}

\section{Detection objective}
\label{sec:methods}
We consider the setting where we observe an undirected simple graph $G=(V,E)$ on $N=|V|$ nodes with $m=|E|$ edges (Fig.~\ref{fig:fig1}(a)) and seek to infer a latent hypergraph $H$ that provides a parsimonious representation of the pairwise edges in $G$. According to the MDL principle~\cite{rissanen1978modeling,grunwald2007minimum,tiago2015model,tiago2025reconstruct,alec2025backbone}, the candidate hypergraph $H$ that should be preferred is the one that gives the shortest two-part description of $G$, which consists of the bits needed to specify $H$ plus the bits needed to specify $G$ given $H$. We first write down the description length of $G$ with no higher-order structure, then the description length of $G$ under a hypergraph, and finally show that their difference decomposes into a gain minus a cost for each hyperedge, which defines both the objective we optimize and a detectability boundary for each candidate hyperedge. Throughout, code lengths use base-2 logarithms and are measured in bits.

\subsection{Description lengths}
\label{sec:baseline}
\label{sec:structural}
\label{sec:attribution}
With no higher-order structure, the simplest description of $G$ is an Erd\H{o}s--R\'enyi code in which each of the $M=\binom{N}{2}$ node pairs has an edge added independently at random with probability equal to the observed density $\rho=m/M$. Writing $Y_{ij}\in\{0,1\}$ for the indicator variable of $(i,j)\in E$, the likelihood for $G$ under this process is $P_{\mathrm{base}}(G\mid\rho)=\rho^m(1-\rho)^{M-m}$, and the baseline description length ignoring higher-order structure can be written as
\begin{equation}
    \label{eq:main_Lbase_def}
    L_{\mathrm{base}}(G\mid \rho) = -m\log\rho-(M-m)\log(1-\rho).
\end{equation}
Every hypergraph description $H$ will be measured against this reference, such that a latent interaction $h\in H$ is worth encoding only if it shortens Eq.~\eqref{eq:main_Lbase_def} by more than the interaction $h$ costs to describe. There are numerous alternative pairwise encodings that can be used in this case, including those that exploit community structure~\cite{peixoto2017nonparametric}, core-periphery~\cite{gallagher2021core} or hub structure~\cite{kirkley2024hubs}. However, these alternative encodings do not result in a direct interpretation of latent higher-order edges in terms of local edge density fluctuations and also do not decouple across the latent hypergraph to enable scalable inference.

A hypergraph explanation $H$ can be used to transmit $G$ in two parts, using first a \emph{model code} that specifies the set of retained hyperedges $H$ and then a \emph{data code} that specifies the edges of $G$ given $H$, thus
\begin{equation}
    \label{eq:two_part}
    L(G,H)=L_{\mathrm{model}}(H)+L_{\mathrm{data}}(G\mid H).
\end{equation}

\emph{Model code}. A hyperedge $h$ with $s=|h|\ge3$ nodes can be specified first by its size and then its node set, through $L_{\mathrm{model}}(h)=L_{\mathrm{size}}(s)+L_{\mathrm{nodes}}(h\mid s)$. For the size we use a prior obtained by integrating the shifted geometric family $P(s\mid\eta)=\eta(1-\eta)^{s-2}$, $s=2,3,\ldots$, over a uniform prior on $\eta$, which gives
\begin{equation}
    \int_0^1\eta(1-\eta)^{s-2}\,d\eta=\frac{1}{s(s-1)}.
\end{equation}
Truncating to $s\le N$ and normalizing with $\sum_{s=2}^{N}[s(s-1)]^{-1}=(N-1)/N$ yields $P(|h|=s)=N/[s(s-1)(N-1)]$ and hence
\begin{equation}
    \label{eq:size_code}
    L_{\mathrm{size}}(s)
    =-\log P(|h|=s)
    =\log\frac{s(s-1)(N-1)}{N}.
\end{equation}
(We keep the normalization over $s\ge2$ for simplicity, since renormalizing over $s\ge3$ lowers every size cost by roughly one bit and has a negligible effect on the results below.) Given its size, the node set can be encoded uniformly among all $s$-subsets of $V$, thus
\begin{equation}
    \label{eq:node_set_code}
    L_{\mathrm{nodes}}(h\mid s)=\log\binom{N}{s},
\end{equation}
so that
\begin{equation}
    \label{eq:Lmodel_h}
    L_{\mathrm{model}}(h)=\log\frac{s(s-1)(N-1)}{N}+\log\binom{N}{s}.
\end{equation}
The model code of a hypergraph is then the sum over its hyperedge-level codes,
\begin{equation}
    \label{eq:Lmodel_H}
    L_{\mathrm{model}}(H)=\sum_{h\in H}L_{\mathrm{model}}(h),
\end{equation}
which requires no separate code for the number or ordering of hyperedges because it is, up to a constant of less than one bit, the code length of an independent-inclusion prior over hyperedge sets (see Appendix~\ref{app:MAP}).

\emph{Data code.} Given $H$, we can encode $G$ by attributing each node pair to either a hyperedge in $H$ or an edge that forms the \emph{pairwise background} of $G$. Because hyperedges may overlap, a node pair may lie inside several of them, so we introduce an \emph{attribution} (or \emph{ownership}) assignment $z$ that maps each pair $(i,j)$ either to the background, $z_{ij}=0$, or to exactly one hyperedge containing it, $z_{ij}=h\in H$, denoting the pairs with $z_{ij}=h$ as the \emph{support} owned by $h$. Background pairs are encoded as $\mathrm{Bernoulli}(\rho)$ with the baseline density, and pairs owned by $h$ as $\mathrm{Bernoulli}(\theta_h)$ with a local density $\theta_h$. Let $T_h$ be the number of pairs owned by $h$, $e_h$ the number of those that are observed edges, and $\hat p_h=e_h/T_h$ their density. Profiling the negative log-likelihood over $\{\theta_h\}$ sets $\theta_h=\hat p_h$ and gives (Appendix~\ref{app:hypergraph-conditioned-data-code})
\begin{equation}
    \label{eq:main_exact_final_identity}
    L_{\rm data}(G\mid H,z,\rho)=L_{\rm base}-\textstyle\sum_{h}T_h\,D_{\rm KL}(\hat p_h\|\rho),
\end{equation}
where $D_{\mathrm{KL}}(a\|b)=a\log(a/b)+(1-a)\log[(1-a)/(1-b)]$ is the Bernoulli KL divergence. Equation~\eqref{eq:main_exact_final_identity} is exact, since the data code equals the baseline code minus, for each hyperedge, the number of bits saved by encoding its owned pairs at their own density rather than the background density. Transmitting the $\theta_h$ themselves would add $O(\log T_h)=O(\log |h|)$ bits per hyperedge, which is negligible against the $\Theta(|h|\log(N/|h|))$ model cost of Eq.~\eqref{eq:Lmodel_h} so is omitted. Likewise, we omit the $O(\log M)$ bits needed to transmit $\rho$ (equivalently $m$), which are common to every hypothesis and therefore cancel from all comparisons. The attribution $z$ is also not transmitted, as it is determined directly by $H$ when no node pair lies in more than one hyperedge, while for overlapping hyperedges $z$ is treated as a latent variable that is optimized jointly with $H$ rather than encoded.

Substituting Eqs.~\eqref{eq:Lmodel_H} and \eqref{eq:main_exact_final_identity} into Eq.~\eqref{eq:two_part}, the total description length of $G$ under $(H,z)$ is
\begin{equation}
    \label{eq:total_exact}
    L(G,H,z)=L_{\mathrm{base}}+\sum_{h\in H}\Bigl[L_{\mathrm{model}}(h)-T_hD_{\mathrm{KL}}(\hat p_h\|\rho)\Bigr].
\end{equation}
Since $L_{\mathrm{base}}$ is fixed by the data $G$, each hyperedge $h$ contributes a cost (its model code) minus a gain (the compression it affords), and substituting in $H=\varnothing$ to Eq.~\eqref{eq:total_exact} recovers the baseline of Eq.~\eqref{eq:main_Lbase_def}.

\subsection{Gain, cost, and detectability}
\label{sec:objective}
Equation~\eqref{eq:total_exact} provides a complete MDL objective for reconstruction over latent hypergraphs $H$, but evaluating it requires an ownership label for all $\binom{N}{2}$ pairs, including the vast majority that are unobserved. For scalability, we track ownership only for observed edges and account for absent pairs inside a hyperedge in aggregate. Let $P_h=\binom{|h|}{2}$ be the number of pairs induced by the pairwise projection of $h\in H$, $\tilde p_h=e_h/P_h$ the density of observed edges over that full pair space, and define
\begin{equation}
    \label{eq:gain_def}
    \mathrm{gain}(h;z)=P_hD_{\mathrm{KL}}(\tilde p_h\|\rho).
\end{equation}
When hyperedges do not overlap, $T_h=P_h$ and Eq.~\eqref{eq:gain_def} equals the exact gain in Eq.~\eqref{eq:total_exact}. When they do overlap, $T_h\le P_h$, and because $n\mapsto nD_{\mathrm{KL}}(e_h/n\|\rho)$ is non-increasing for $e_h/n\ge\rho$ (Appendix~\ref{app:hypergraph-conditioned-data-code}), every candidate with $\tilde p_h\ge\rho$ satisfies $\mathrm{gain}(h;z)\le T_hD_{\mathrm{KL}}(\hat p_h\|\rho)$. The objective we minimize, given by
\begin{align}
    \label{eq:additive_total_main}
    L(G,H,z)
    &=
    L_{\mathrm{base}}+
    \sum_{h\in H}
    \left[
    L_{\mathrm{model}}(h)-\mathrm{gain}(h;z)
    \right],
\end{align}
is therefore an upper bound on the description length of Eq.~\eqref{eq:total_exact}, which is tight in the non-overlapping case. Eq.~\eqref{eq:additive_total_main} provides an advantageous representation because the gain is computed over all $P_h$ internal pairs of $h$, disfavoring hyperedges whose owned edges cover only a small fraction of its pairwise projection. This discourages redundant overlapping hyperedges. We refer to Eq.~\eqref{eq:additive_total_main} as the description length throughout (Fig.~\ref{fig:fig1}(b)). Up to these approximations, minimizing this objective corresponds to a penalized profile-likelihood under the independent-inclusion prior of Appendix~\ref{app:MAP}.

For fixed $P_h$, the gain is minimized at $\tilde p_h=\rho$ and increases as $\tilde p_h$ moves away from $\rho$ in either direction, whereas a latent interaction should project to support that is locally \emph{denser} than the background. Thus, the search of Sec.~\ref{sec:search} rejects candidates with $\tilde p_h<\rho$ by convention. (For sparse networks $\rho$ is small and this condition is rarely invoked.) Additionally, a hyperedge $h$ lowers the objective of Eq.~\eqref{eq:additive_total_main} only when its gain exceeds its cost, which we summarize by the gain--cost ratio
\begin{equation}
    \lambda_h
    =
    \frac{\mathrm{gain}(h;z)}
    {L_{\mathrm{model}}(h)}
    =
    \frac{P_hD_{\mathrm{KL}}(\tilde p_h\Vert\rho)}
    {L_{\mathrm{model}}(h)},
    \label{eq:lambda_def}
\end{equation}
so that $\lambda_h=1$ provides a hyperedge-level detectability boundary induced by the objective.

\begin{algorithm*}[t]
    \caption{Seed-centered merge--reassign--split--reassign search}
    \label{alg:search}
    \KwIn{Observed graph $G=(V,E)$}
    \KwOut{Reconstructed hypergraph $\hat H$}
    Initialize $\hat H\,=\varnothing$ (the pairwise baseline) and assign all observed edges to the background.\\
    \Repeat{no stage lowers $L$ by more than the tolerance for two consecutive iterations}{
        \quad Select merge seeds\;
        \quad Generate merges by single-node expansion, beam-search expansion, and merging with nearby retained hyperedges\;
        \quad Evaluate candidate merges using $\Delta L$ and greedily accept improving, nonconflicting moves\;
        \quad Reassign observed edges in the affected region\;
        \quad Select split seeds\;
        \quad Generate splits by single-node and beam-search pruning, deletion to pairwise baseline, and bipartition splitting\;
        \quad Evaluate candidate splits using $\Delta L$ and greedily accept improving, nonconflicting moves\;
        \quad Reassign observed edges in the affected region\;
    }
    \Return{$\hat H$}\;
\end{algorithm*}

\begin{figure*}[t]
    \centering
    \includegraphics[width=12cm]{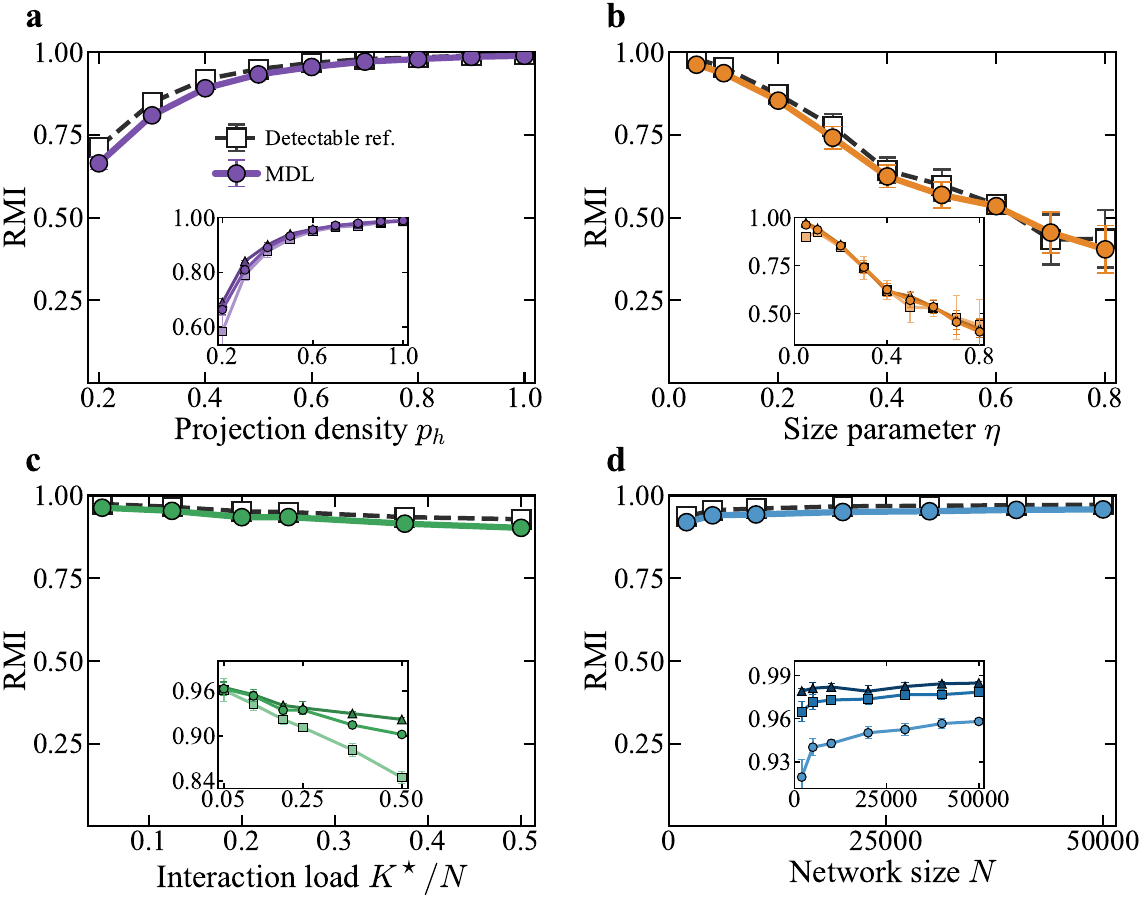}
    \caption{\textbf{Reconstruction performance and detectability on synthetic higher-order networks.}
    (a) Reconstruction accuracy versus the common projection density $p_h$ shared by all planted interactions, with $N=5000$, $K^\star/N=0.2$ and $\eta=0.1$. Solid curves with filled markers show the MDL reconstruction, while dashed curves with open markers show the detectable reference $H_{+}^{\star}$ (the planted interactions with $\lambda_h>1$, Sec.~\ref{sec:interaction_recovery}). The insets in (a)--(c) repeat the same experiment for $N\in\{2000,5000,10000\}$ (squares, circles, and triangles, respectively).
    (b) Reconstruction accuracy versus size parameter $\eta$ (expected interaction size $\mathbb{E}[s]=1+\eta^{-1}$), with $K^\star/N=0.2$ and $p_h\sim U(0.2,1.0)$.
    (c) Reconstruction accuracy versus interaction load $K^\star/N$, with $\eta=0.1$ and $p_h\sim U(0.2,1.0)$. Main curves in (a)--(c) use $N=5000$.
    (d) Reconstruction accuracy versus number of nodes $N$ for $p_h\sim U(0.2,1.0)$. The inset compares circles, squares, and triangles denoting $p_h\sim U(0.2,1.0)$, $U(0.5,1.0)$, and $U(0.7,1.0)$, respectively, with $K^\star/N=0.2$ and $\eta=0.1$. Each panel reports the reconstruction accuracy (using RMI~\cite{jerdee2025RMI}) as the mean $\pm$ three standard errors over five independent realizations, with both $H^\star$ and the pairwise projection independently redrawn in each realization. Reconstructions use the default algorithmic settings of Appendix~\ref{app:defaults}.
    }
    \label{fig:fig2}
\end{figure*}

Let $e(s)$ denote the number of observed edges attributed to a representative hyperedge candidate of size $s$, and define $\tilde p(s)=e(s)/P_s$, where $P_s=\binom{s}{2}$. Since $L_{\mathrm{model}}(h)$ depends on $h$ only through $|h|$, we write $L_{\mathrm{model}}(s)$ for its common value over all $h$ with $|h|=s$. Retention of $h$ in the inferred hypergraph $H$ requires $P_sD_{\mathrm{KL}}(\tilde p(s)\Vert\rho)>L_{\mathrm{model}}(s)$. For $s=o(N)$, Stirling's approximation gives 
\begin{align}
    \log\binom{N}{s}&=s\log\frac{N}{s}+s\log e+O\!\left(\log s+\frac{s^2}{N}\right),\\
    L_{\mathrm{model}}(s)&=\log\binom{N}{s}+L_{\rm size}(s)\\
    &=s\log\frac{N}{s}+O(s+\log s),
\end{align}
where the second line uses $L_{\rm size}(s)=O(\log s)$. Dividing this condition by $s$ gives
\begin{equation}
    \frac{s-1}{2}
    D_{\mathrm{KL}}\!\left(\tilde p(s)\Vert\rho\right)
    \gtrsim
    \log\frac{N}{s}
    \label{eq:scaling_retention_full}
\end{equation}
for detectability of a latent hyperedge $h$ of size $s=|h|$ against the background of density $\rho$. Consider the sparse-background regime $\rho=\gamma/N$, with $\gamma=\Theta(1)$, and suppose that a constant fraction $c\in(0,1]$, independent of $N$, of the induced pair space is observed, so that $\tilde p(s)=c$ and $e(s)=\Theta(s^2)$. Then, for $c$ bounded away from zero, $D_{\mathrm{KL}}\!\left(c\middle\|\frac{\gamma}{N}\right)=c\log N+O(1)$, and Eq.~\eqref{eq:scaling_retention_full} becomes
\begin{equation*}
    \frac{s-1}{2}\bigl[c\log N+O(1)\bigr]\gtrsim\log\frac{N}{s}.
\end{equation*}
Since $\log(N/s)=\log N-\log s$, for bounded $s$ the difference between the leading evidence and cost terms is $[c(s-1)/2-1]\log N$. Thus, the leading-order MDL detectability condition is
\begin{equation}
    c(s-1)>2.
    \label{eq:leading_detectability_boundary}
\end{equation}
This exposes a simple size vs density tradeoff, in which lower internal density $c$ can be compensated by a larger interaction size $s$. (Complementary theoretical work has established information theoretic thresholds for recovering random uniform hypergraphs from projections~\cite{bresler2024thresholds,bresler2025partial,gong2026detection}, but these do not address non-uniform hypergraphs.) Equivalently, the leading-order detectability boundary is governed by the mean internal degree $c(s-1)$, with candidate retention favored once this quantity exceeds two. For fixed interaction size $s$, both the gain and the specification cost grow as $\log N$, so increasing $N$ alone does not shift the leading-order boundary. Eq.~\eqref{eq:leading_detectability_boundary} holds for bounded $s$, and the complementary regime $s\to\infty$ is treated in Sec.~\ref{sec:information_boundary}. (The two results are separate asymptotic statements about the same criterion, $\lambda_h=1$.) The exact ratio $\lambda_h$ is used in the experiments of Sec.~\ref{sec:interaction_recovery}. For the smallest interactions, the $O(1)$ corrections are relevant, since at $s = 3$, Eq.~\eqref{eq:leading_detectability_boundary} cannot be satisfied for any $c\le1$, and even the exact criterion gives $\lambda_h < 1$ for a fully observed triangle whenever the mean degree $\rho(N-1)$ exceeds $[(N-1)/(N-2)]^{1/3}\approx 1$. Three-node hyperedges are therefore retained only when the background is extremely sparse, and the method should be understood as targeting hyperedges of size four and above in typical sparse regimes (this is the origin of the decline in recovery at large $\eta$ in Fig.~\ref{fig:fig2}(b)).

\subsection{Optimality of the detectability boundary}
\label{sec:information_boundary}
To ask whether the detectability boundary reflects a statistical limit on recovering higher-order structure from pairwise observations, we consider a canonical single-interaction setting with either no hyperedge or a single hyperedge of known size $s$ over an unknown node set $S$. Under the background model, which here we denote with $\mathbb P_0$, all pair indicators are independent ${\rm Bernoulli}(\rho)$. Meanwhile, under the planted model, which we denote with $\mathbb P_S$, pairs internal to $S$ are distributed as ${\rm Bernoulli}(c)$ and all remaining pairs are distributed as ${\rm Bernoulli}(\rho)$, with $0<\rho<c<1$. The total information separating a planted interaction from the background is therefore
\begin{equation}
    I_s:=D_{\rm KL}(\mathbb P_S\Vert\mathbb P_0)=P_s\,D_{\rm KL}(c\Vert\rho),
\end{equation}
where $P_s=\binom{s}{2}$. Reliable reconstruction requires both vanishing false detection under $\mathbb P_0$ and asymptotically vanishing normalized localization error under $\mathbb P_S$.

Consider $N\to\infty$ with $s=s_N\to\infty$ and $s=o(N)$. Appendix~\ref{app:information_boundary} establishes that, for every fixed $\epsilon>0$, reliable reconstruction is impossible if $I_s<(1-\epsilon)\log\binom Ns$, whereas it is achievable if $I_s>(1+\epsilon)\log\binom Ns$. The first implication holds for every estimator, whereas the second can be attained by a thresholded maximum-likelihood scan under mild regularity conditions, which are satisfied, for example, when $c$ stays bounded away from $0$ and $1$ and $c-\rho$ remains bounded below by a positive constant. Hence, we have that
\begin{equation}
    I_s\sim\log\binom Ns
    \label{eq:sharp_information_scale}
\end{equation}
is the sharp leading information scale for reconstruction of an individual latent hyperedge of size $s$. 

In the same regime ($s\to\infty$, $s=o(N)$), the structural cost satisfies
\begin{equation}
    \begin{aligned}
    L_{\rm model}(s)&=\log\binom Ns +\log\frac{s(s-1)(N-1)}{N}\\
    &=[1+o(1)]\log\binom Ns,
    \end{aligned}
    \label{eq:model_cost_asymptotic}
\end{equation}
so that Eqs.~\eqref{eq:sharp_information_scale} and~\eqref{eq:model_cost_asymptotic} give
\begin{equation}
    \frac{I_s}{L_{\rm model}(s)}\to1
\end{equation}
at the sharp reconstruction boundary. Thus the statistical limit derived independently of the MDL code coincides, at leading order, with the unit gain--cost boundary. This single interaction problem is the classical planted dense subgraph problem, for which information-theoretic detection and recovery thresholds of the same $\log\binom{N}{s}$ scale are known~\cite{butucea2013detection,ariascastro2014community,hajek2017information}, although the threshold for simply detecting the presence of a planted subgraph can differ from the recovery threshold in some regimes. Because the derived criterion requires localizing $S$, the boundary derived here is a recovery (reconstruction) boundary, and we use ``detectability'' in this sense throughout. The key result in Eq.~\eqref{eq:model_cost_asymptotic} is that the MDL specification cost of a hyperedge falls exactly at this scale. Appendix~\ref{app:information_boundary} further shows that, for known $s$, the exact (exhaustive) minimizer of the description length objective (which, unlike the likelihood scan, requires no knowledge of $c$) achieves reliable reconstruction above this boundary. With the converse below the boundary, this shows that the proposed description length objective attains the statistically optimal leading boundary in the single interaction setting. (Exhaustive minimization is computationally infeasible, and we make no claim that the local search of Sec.~\ref{sec:search} attains this boundary in general.) While the full reconstruction problem is more general, allowing multiple interactions to coexist, the same hyperedge-level gain--cost quantity $\lambda_h$ remains directly computable for each interaction, and Sec.~\ref{sec:interaction_recovery} shows that it sharply predicts recovery in the full hypergraph setting.

\section{Reconstruction algorithm}
\label{sec:search}
In our reconstruction problem we aim to find a hypergraph $H$ and an ownership assignment $z$ that minimize the objective in Eq.~\eqref{eq:additive_total_main}. Since $G$ and $\rho$ are fixed, the baseline term is constant with respect to $(H,z)$, and the optimization reduces to the sum of contributions from retained hyperedges,
\begin{equation}
    \label{eq:additive_optimization}
    \min_{\{H,z\}}\left\{\sum_{h\in H}\left[L_{\mathrm{model}}(h)-\mathrm{gain}(h;z)\right]\right\}.
\end{equation}
We optimize this objective with a seed-centered local merge--reassign--split--reassign search that relies on two properties. The first is that latent interactions produce locally concentrated pairwise support, so candidates can be found by local expansion and refinement rather than global enumeration. The second is that the additive form of Eq.~\eqref{eq:additive_optimization} implies that a proposed update changes the objective only through hyperedges that are created, removed, or whose owned support changes, so $\Delta L$ can be computed from the affected terms alone (Appendix~\ref{sec:local_evaluation}). These two forms of locality, in the search space and in computation, allow moves from many seeds to be generated and scored in batches.

\subsection{Optimization strategy}
\label{sec:seeds}
The proposed search algorithm repeats the cycle $\text{merge}\to\text{reassign}\to\text{split}\to\text{reassign}$ (Fig.~\ref{fig:fig1}(c)), starting from the pairwise baseline on $E$ ($H=\varnothing$). Here, a seed is a selected observed pair or retained hyperedge that serves as the starting point for local candidate generation. Merge moves grow locally supported candidates from pair or hyperedge seeds by adding nodes or merging nearby retained hyperedges, while split moves contract, partition, or remove hyperedges whose size is no longer justified by their gain. Each seed generates candidates only by local expansion or refinement, and only a selected subset of seeds is explored at each stage, since nearby seeds induce strongly overlapping candidate neighborhoods. Pair seeds initiate new higher-order candidates, while hyperedge seeds refine existing ones. Pair seeds are ranked by common-neighbor statistics of their endpoints and hyperedge seeds by recent activity, because an accepted edit changes gains only within its affected neighborhood, so hyperedges far from recent edits tend to reproduce previously explored moves. We also reserve a random fraction of each seed budget to preserve exploration (Appendix~\ref{app:search_algorithm}).

Candidate moves from multiple selected seeds are generated and scored simultaneously within the same proposal batch. Each move is scored by its local $\Delta L$, and a set of improving, nonconflicting moves is accepted greedily in each batch. Reassignment then reconsiders observed edges in the affected region over their admissible owners and changes ownership only when doing so decreases the objective. Because ownership enters the gain directly, reassignment can decrease the supported pairwise edge set of redundant hyperedges, exposing them to later split moves. The search terminates when two consecutive outer iterations fail to improve the objective beyond a tolerance set by absolute and relative thresholds (Appendix~\ref{app:search_algorithm}). (Fig.~\ref{fig:fig1}(d) shows a representative trajectory in which the description length $L$ decreases monotonically while reconstruction accuracy saturates well before termination.) The search procedure is summarized in Algorithm~\ref{alg:search}, in which $\hat H$ denotes the set of retained higher-order hyperedges, and the displayed reconstruction consists of $\hat H$ together with the observed edges that remain in the pairwise background. Further implementation details are provided in Appendix~\ref{app:search_algorithm}, and the full implementation is available at
\url{https://github.com/zangyingbang/KLMDL-Hypergraph-Inference}.

\begin{figure}[t]
    \centering
    \includegraphics[width=0.8\columnwidth]{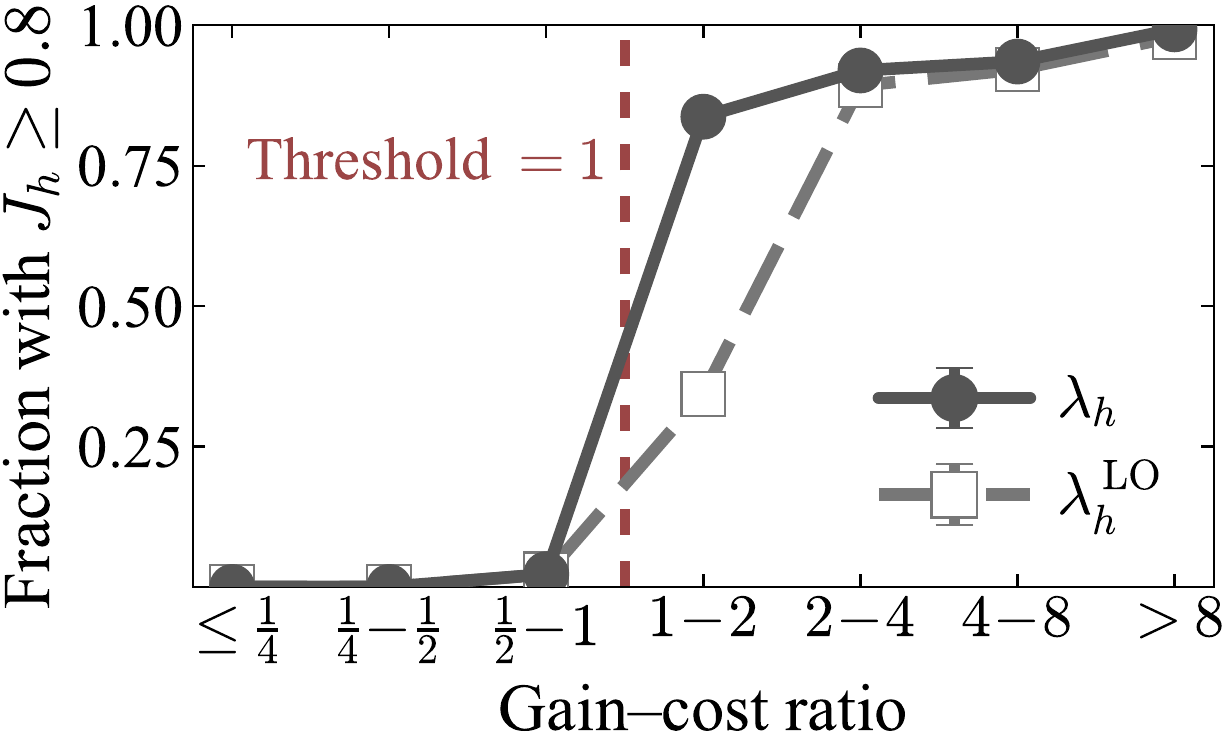}
    \caption{
    \textbf{Interaction-level recovery against the detectability threshold.} Recovery accuracy of interactions is disaggregated across measured intervals of the gain--cost ratio in $\lambda_h$ (Eq.~\eqref{eq:lambda_def}) and its leading-order approximation $\lambda_h^{\rm LO}$ (Eq.~\eqref{eq:lambda_lo}) for all four controlled synthetic experimental scans of Fig.~\ref{fig:fig2}. For each quantity and realization, planted hyperedges are grouped into the given intervals along the $x$-axis and the fraction attaining Jaccard overlap $J_h\geq0.8$ with the ground truth is computed within each interval, with $\lambda_h$ and $\lambda_h^{\rm LO}$ evaluated on the planted node sets as described in Sec.~\ref{sec:interaction_recovery}. Points and error bars indicate the mean and three standard errors across the five simulations, and the dashed vertical line separates intervals below and above the interaction-level detectability boundary.}
    \label{fig:fig3}
\end{figure}

\subsection{Computational scaling}
\label{sec:scaling_theory}
\begin{figure*}[t]
    \centering
    \includegraphics[width=12cm]{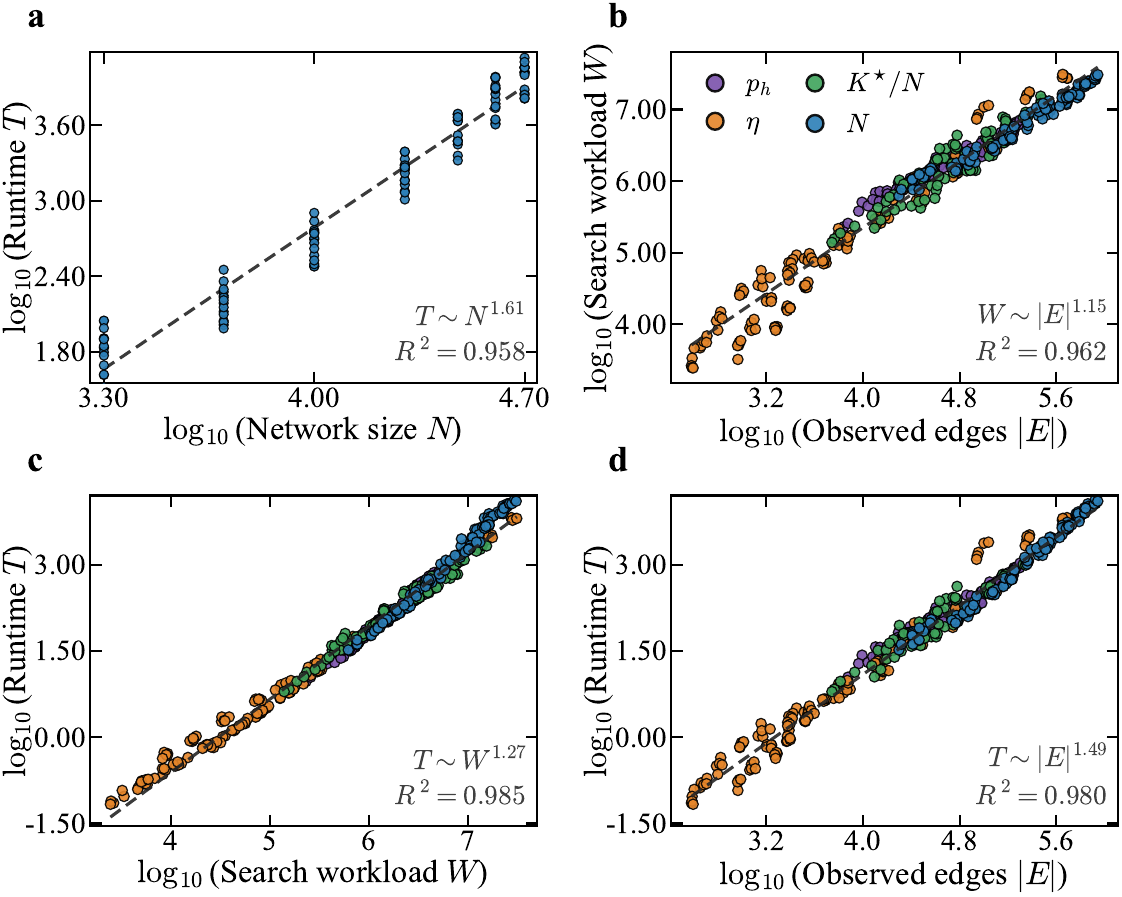}
    \caption{\textbf{Computational scaling of the reconstruction algorithm.} 
    (a)~Runtime versus network size $N$, pooling all simulations of Fig.~\ref{fig:fig2}(d).
    (b) Search workload $W$ versus number of observed edges $|E|$, over all simulations from Fig.~\ref{fig:fig2}, with the different parameter sweeps disaggregated by color. Dashed lines are ordinary least-squares fits on the log-log scale, whose slopes give the nonlinear scaling exponents. (c)-(d) Runtime versus search workload and number of edges, for the same set of simulations.}
    \label{fig:fig4}
\end{figure*}

At outer iteration $t$ (e.g. the $t$-th $\text{merge}\to\text{reassign}\to\text{split}\to\text{reassign}$ cycle), let $M_t$ and $S_t$ denote the numbers of merge and split candidate states explored by the local search, and let $A_t\subseteq E$ denote the set of observed edges visited during reassignment of the affected regions. We estimate the total number of computations done locally as
\begin{equation}
    W=\sum_{t=1}^{\mathcal{T}}\left(M_t+S_t+|A_t|\right),
    \label{eq:workload}
\end{equation}
where $\mathcal{T}$ is the number of outer iterations. If each part of the observed pair layer participates in only a bounded amount of local exploration and reassignment, the total computational workload scales linearly with the observed edge count $m=\abs{E}$. However, in practice, seed-centered regions may overlap, accepted edits may trigger nearby reassignment, and local structures may be revisited before the algorithm converges. We can absorb these effects into a \emph{revisitation factor} $\kappa$, giving the scaling estimate
\begin{equation}
    W=O(\kappa |E|).
\end{equation}
The regime $\kappa=O(1)$ therefore corresponds to linear scaling in the observed edge layer, while larger $\kappa$ captures additional work caused by overlap and repeated refinement (Sec.~\ref{sec:scaling_results}). We find only slightly superlinear scaling of the workload in practice (Fig.~\ref{fig:fig4}), such that $\kappa\propto|E|^{0.15}$, so that the algorithm is highly scalable to large empirical systems.

\section{Results}
\label{sec:results}

\subsection{Reconstruction of planted higher-order structure}
\label{sec:synthetic}
We first evaluate our method on synthetic networks generated by a two-stage process that instantiates the observation model of Fig.~\ref{fig:fig1}(a) by planting a latent hypergraph $H^\star$ and passing only its noisy pairwise projection $G=(V,E)$ to the algorithm.

To plant the latent hypergraph, we fix a node set $V$ with $|V|=N$ and a number of planted interactions $K^\star$. For each $k=1,\ldots,K^\star$, we independently draw $s_k$ from $P(s\mid\eta)=\eta(1-\eta)^{s-2},\ s=2,3,\ldots$, truncated to $s\le N$, and then draw the node set $h_k\subseteq V$ uniformly at random among the $\binom{N}{s_k}$ subsets of that size. The size parameter $\eta\in(0,1)$ therefore controls the interaction size distribution of the benchmark through $\mathbb{E}[s_k]=2+(1-\eta)/\eta=1+1/\eta$, so that small $\eta$ produces large interactions and $\eta\to1$ reduces to a primarily dyadic system. The latent hyperedges are drawn independently and may overlap arbitrarily in their node sets, so the normalized interaction load $K^\star/N$ controls how much overlap the pairwise projection contains (and hence how noisy the attribution problem becomes). To project to pairwise observations, for each planted interaction $h_k$ we draw a projection density $p_{h_k}\sim U(a,b)$ independently and then retain each of the $\binom{s_k}{2}$ internal pairs of $h_k$ independently with probability $p_{h_k}$. The observed graph $G$ is the simple graph on $V$ whose edge set $E$ is the union of all retained pairs over all interactions. A pair retained by more than one interaction appears only once, making overlapping interactions harder to distinguish. Drawing $p_{h}$ per interaction rather than globally means that a single problem instance contains both well observed and poorly observed interactions. In the projection-density sweep of Fig.~\ref{fig:fig2}(a), we instead use a common $p_h$ shared by all planted interactions and vary this value along the horizontal axis. Planted interactions of size two coincide with single observed edges and become part of the background in the ground truth labeling (see Appendix~\ref{app:rmi} for more details). All counts of planted higher-order interactions below refer to $|h_k|\ge3$, whereas $K^\star$ counts all planted interactions (including dyads).

\begin{table*}[t]
    \caption{\textbf{Analysis of empirical hypergraphs.} The observed pairwise graph $G$ is generated from the recorded hyperedges $H^\star$ by retaining each internal pair of each hyperedge $h\in H^\star$ independently with probability $p_h\sim U(0.5,1)$. $N$ and $|E|$ are the node and pairwise edge counts, and $\rho=|E|/\binom{N}{2}$ the pairwise graph's density. RMI is measured between the inferred and ground truth hyperedges, $s_L=(L_{\rm base}-L)/L_{\rm base}$ is the relative description-length saving over the pairwise baseline, and $T$ is wall-clock runtime~\cite{runtime_note}. Each row is the result of a single run under the default configuration (Appendix~\ref{app:defaults}).}
    \label{tab:empirical_results}
    \centering
    \begin{ruledtabular}
    \begin{tabular}{lrrrrrr}
    Dataset & $N$ & $|E|$ & $\rho$ & RMI & $s_L(\%)$ & $T$ \\
    \hline
    Languages & $610$ & $5{,}909$ & $0.0318$ & $0.861$ & $47.3$ & $20.8\,\mathrm{s}$ \\
    Foursquare & $2{,}861$ & $56{,}539$ & $0.0138$ & $0.865$ & $42.1$ & $10.7\,\mathrm{min}$ \\
    Board of Directors & $1{,}013$ & $2{,}275$ & $0.0044$ & $0.785$ & $31.2$ & $1.1\,\mathrm{s}$ \\
    Terrorism & $230$ & $1{,}938$ & $0.0736$ & $0.751$ & $38.9$ & $3.1\,\mathrm{s}$ \\
    Pollination & $678$ & $23{,}935$ & $0.1043$ & $0.923$ & $41.9$ & $3.4\,\mathrm{min}$ \\
    American Revolution & $136$ & $2{,}340$ & $0.2549$ & $0.921$ & $19.8$ & $3.9\,\mathrm{s}$ \\
    DBLP 1980 & $15{,}591$ & $131{,}825$ & $0.0011$ & $0.976$ & $66.1$ & $2.6\,\mathrm{min}$ \\
    DBLP 1985 & $27{,}202$ & $245{,}607$ & $0.0007$ & $0.968$ & $66.6$ & $11.3\,\mathrm{min}$ \\
    DBLP 1990 & $57{,}247$ & $518{,}454$ & $0.0003$ & $0.969$ & $67.4$ & $54.9\,\mathrm{min}$ \\
    \end{tabular}
    \end{ruledtabular}
\end{table*}

A benchmark setting is thus specified by the parameter configuration $(N,K^\star,\eta,p_h)$, with the particular parameter specifications given in the caption of Fig.~\ref{fig:fig2}. Reconstruction quality is measured by the reduced mutual information (RMI)~\cite{jerdee2025RMI} (Appendix~\ref{app:rmi}), which equals one when the inferred and ground-truth edge labelings coincide up to relabeling. We find that the method achieves high recovery over a broad range of settings, with the insets showing that the same qualitative behavior persists across a range of network sizes. Across all panels, we can see that recovery performance decreases when the pairwise projection provides weaker or more ambiguous evidence for the latent interactions. In Fig.~\ref{fig:fig2}(a), incomplete observations at low $p_h$ reduce the observed support within each hyperedge. In Fig.~\ref{fig:fig2}(b) smaller interactions reduce the pair space $P_h$. And in Fig.~\ref{fig:fig2}(c) larger interaction loads $K^{\star}/N$ create greater overlap among the projected hyperedges, which makes latent higher-order structure harder to distinguish from the background. In Fig.~\ref{fig:fig2}(d), recovery remains consistently high as $N$ increases, with RMI values all exceeding 0.90. Under this setting with $p_h\sim U(0.2,1.0)$, networks with $N=50{,}000$, $K^{\star}=10{,}000$, and approximately $5.94\times10^5$ observed edges are reconstructed with mean $\mathrm{RMI}=0.958$ and average runtime $1.92$~h over five runs~\cite{runtime_note}, without imposing any maximum interaction order.

\begin{table*}[htb]
    \caption{\textbf{Comparison with Bayesian reconstruction on synthetic data.} We compare the proposed reconstruction method with that of Ref.~\cite{young2021reconstruct}. The synthetic experiments follow the same settings as in Fig.~\ref{fig:fig2}(d), with $p_h\sim U(0.2,1.0)$, $\eta=0.1$, and $K^{\star}=0.2N$, with $N$ restricted to smaller sizes so both methods can be run. RMI values are means $\pm$ one standard error, and runtimes are reported as the medians over the simulations, with the range shown in brackets. The speedup is the ratio of the median runtimes.}
    \label{tab:bayesian_comparison}
        \begin{ruledtabular}
            \begin{tabular}{ccccccc}
            $N$ &
            $\mathrm{RMI}_{\rm MDL}$ &
            $\mathrm{RMI}_{\rm Bayes}$ &
            $\Delta\mathrm{RMI}$ &
            Runtime (MDL) &
            Runtime (Bayes) &
            Speedup \\
            \hline
            1000 & $0.897 \pm 0.004$ & $0.748 \pm 0.009$ & 0.149 & 34.3 s [15.9 s, 52.3 s] & 16.4 min [47.1 s, 10.35 h] & $28.6\times$ \\
            2000 & $0.930 \pm 0.004$ & $0.772 \pm 0.005$ & 0.157 & 52.6 s [28.7 s, 59.8 s] & 22.3 min [3.0 min, 28.95 h] & $25.4\times$ \\
            3000 & $0.934 \pm 0.002$ & $0.779 \pm 0.004$ & 0.155 & 1.14 min [1.08 min, 1.36 min] & 4.66 h [16.2 min, 59.64 h] & $245.0\times$ \\
            4000 & $0.933 \pm 0.003$ & $0.781 \pm 0.004$ & 0.152 & 1.38 min [1.05 min, 1.76 min] & 1.31 h [1.26 h, 7.58 h] & $57.1\times$ \\
            5000 & $0.937 \pm 0.001$ & $0.779 \pm 0.004$ & 0.158 & 2.22 min [1.69 min, 2.66 min] & 6.93 h [45.9 min, 16.53 h] & $187.5\times$ \\
        \end{tabular}
    \end{ruledtabular}
\end{table*}

\subsection{Detectability and computational scaling in synthetic benchmarks}
\label{sec:interaction_recovery}
\label{sec:scaling_results}
The detectability boundary of Sec.~\ref{sec:objective} also can be used to predict which planted interactions can be recovered in these experiments. We can first remove from the ground truth every interaction with $\lambda_h\leq 1$, with $\lambda_h$ defined in Eq.~\eqref{eq:lambda_def}. (For a planted $h$, $e_h$ counts all observed edges among its $P_h$ internal pairs, whether or not they are shared with other planted interactions.) We can then define the network-level detectability reference $H_+^\star$ as this set of detectable hyperedges, and measure the accuracy $\mathrm{RMI}(H_+^\star;H^\star)$ of this reference itself, which serves as an approximate ceiling for the reconstruction performance (open markers in Fig.~\ref{fig:fig2}). We can observe decreases in the same regimes as the reconstruction, with close agreement between the detectable reference and the inferred hypergraph indicating that the local search recovers nearly all of the detectable structure without enumerating the exponentially large candidate space.

\begin{table*}[htb]
    \caption{\textbf{Comparison with Bayesian reconstruction on empirical data.} Both methods are evaluated using the subsampled pairwise projections of the empirical hypergraph datasets described in Table~\ref{tab:empirical_results}. Entries marked ``OOM'' indicate that the Bayesian method exceeded the 80-GB memory limit and terminated without producing a reconstruction.}
    \label{tab:bayesian_empirical}
        \begin{ruledtabular}
            \begin{tabular}{lrrcccc}
            Dataset &
            $N$ & $|E|$ & $\mathrm{RMI}_{\rm MDL}$ & $\mathrm{RMI}_{\rm Bayes}$ & Runtime (MDL) & Runtime (Bayes) \\
            \hline
            Languages       & $610$    & $5{,}909$   & 0.861 & 0.789 & 20.8 s   & 10.5 h \\
            Foursquare      & $2{,}861$& $56{,}539$  & 0.865 & --    & 10.7 min & OOM \\
            Board of Directors   & $1{,}013$& $2{,}275$   & 0.785 & 0.734 & 1.1 s    & 3.5 s \\
            Terrorism       & $230$    & $1{,}938$   & 0.751 & 0.696 & 3.1 s    & 31.1 min \\
            Pollination     & $678$    & $23{,}935$  & 0.923 & --    & 3.4 min  & OOM \\
            American Revolution   & $136$    & $2{,}340$   & 0.921 & 0.783 & 3.9 s    & 3.1 min \\
            DBLP 1980       & $15{,}591$& $131{,}825$& 0.976 & --    & 2.6 min  & OOM \\
            DBLP 1985       & $27{,}202$& $245{,}607$& 0.968 & --    & 11.3 min & OOM \\
            DBLP 1990       & $57{,}247$& $518{,}454$& 0.969 & --    & 54.9 min & OOM \\
            \end{tabular}
        \end{ruledtabular}
\end{table*}

The same results hold at the level of individual planted interactions (Fig.~\ref{fig:fig3}). Applying the leading-order approximation of Eq.~\eqref{eq:leading_detectability_boundary} to a candidate hyperedge $h$ of size $s=|h|$ in the sparse regime, where 
\begin{align}
    \mathrm{gain}(h;z)&=\binom{s}{2}\bigl[\tilde p_h\log N+O(1)\bigr],\\
    L_{\rm model}(h)&=s\log N+O(s\log s),
\end{align}
gives $\lambda_h=\tilde p_h(|h|-1)/2+O(s\log s/\log N)$. We denote the leading-order term of this result with 
\begin{equation}
    \lambda_h^{\rm LO}=\frac{\tilde p_h(|h|-1)}{2}.
    \label{eq:lambda_lo}
\end{equation}
The condition $\lambda_h^{\rm LO}=1$ is precisely the boundary of Eq.~\eqref{eq:leading_detectability_boundary} with $c$ replaced by $\tilde p_h$. We then have that $\lambda_h>1$ and $\lambda_h^{\rm LO}>1$ give, respectively, the full detection criterion and its leading-order approximation. We examine how interaction-level recovery varies with these two quantities and whether their detectability boundaries separate accurately recovered from poorly recovered interactions in practice. To assess this, planted and inferred hyperedges are greedily matched one-to-one in decreasing order of Jaccard overlap $J_h\in [0,1]$, with $J_h=0$ when $h$ is not matched. Here $J_h=1$ indicates exact recovery of the planted node set, while $J_h\geq0.8$ is taken to indicate accurate node set recovery for Fig.~\ref{fig:fig3}. Across all four controlled synthetic experiments in Fig.~\ref{fig:fig2}, the analysis includes $767{,}083$ planted hyperedges. Pooling these interactions, we find that $76.1\%$ of those with $\lambda_h^{\mathrm{LO}}>1$ attain $J_h\geq0.8$, compared with only $1.55\%$ for $\lambda_h^{\mathrm{LO}}\leq1$, so the leading-order approximation captures the broad size--density dependence of practical hyperedge recovery. The full gain--cost ratio provides a sharper separation, with $89.7\%$ of interactions with $\lambda_h>1$ attaining $J_h\geq0.8$ compared with only $1.28\%$ for $\lambda_h\leq1$, and the fraction attaining $J_h\geq0.8$ approaches one as $\lambda_h$ increases (Fig.~\ref{fig:fig3}). These results illustrate the practical value of our detectability limit for predicting the recoverability of latent hyperedges.

Fig.~\ref{fig:fig4}(a) plots the runtime $T$ against the network size $N$ for the simulations of Fig.~\ref{fig:fig2}(d), Fig.~\ref{fig:fig4}(b) plots the search workload $W$ against the number of observed edges $|E|$ over all simulations of Fig.~\ref{fig:fig2}, and Fig.~\ref{fig:fig4}(c,d) plot $T$ against $W$ and $|E|$, respectively, for the same simulations. We find that $T\sim N^{1.61}$ for the simulations of Fig.~\ref{fig:fig2}(d), and $W\sim|E|^{1.15}$ and $T\sim|E|^{1.49}$ over all simulations, with the intermediate relation $T\sim W^{1.27}$ that we attribute to the evaluation cost of each hyperedge growing with its size. The workload exponent is close to the linear behavior expected for $\kappa=O(1)$, so $\kappa$ grows only weakly with $|E|$, and runtime is subquadratic in the number of observed edges.

\subsection{Empirical hypergraphs and comparison with existing methods}
\label{sec:empirical}
\label{sec:comparison}
Table~\ref{tab:empirical_results} reports results on benchmarks built from nine real-world hypergraphs from social, ecological, linguistic, and bibliographic systems, where the recorded hyperedges serve as ground truth and the observed pairwise network $G$ is generated by retaining each internal pair of each hyperedge with probability $p_h\sim U(0.5,1)$, drawn once per hyperedge (Appendix~\ref{app:datasets}). We find that the reconstruction RMI ranges from $0.75$ to $0.98$, and the largest dataset, DBLP 1990 ($N=57{,}247$, $|E|=5.18\times10^5$), is reconstructed in under one hour, consistent with the scaling of Fig.~\ref{fig:fig4}. The description length saving over the pairwise baseline, which reaches $67.4\%$ on DBLP 1990, shows that the higher-order description is also considerably more parsimonious than the pairwise one in these empirical systems.

We compare our method quantitatively against the Bayesian approach of Ref.~\cite{young2021reconstruct} and the motif-cover approach of Ref.~\cite{wegner2024nonparametric} introduced in Sec.~\ref{sec:intro}. The synthetic comparison uses the main setting of Fig.~\ref{fig:fig2}(d), with $p_h\sim U(0.2,1.0)$, $\eta=0.1$, and $K^{\star}=0.2N$. Table~\ref{tab:bayesian_comparison} provides the results of the comparison. We observe that the proposed method achieves consistently higher RMI than the Bayesian method, with an absolute improvement of approximately $0.15$ across all tested sizes. We note that the Bayesian model assumes every latent interaction projects to a complete clique, so the incomplete projections used here (and $p_h\sim U(0.5,1)$ in Table~\ref{tab:bayesian_empirical}) lie outside its model class, whereas our objective makes no such assumption. (Both methods, however, receive identical observations.) The proposed method also has median runtimes that are $25$--$245\times$ faster, with far less runtime variability, and it handles the $N=50{,}000$ instance of Sec.~\ref{sec:synthetic}, which was computationally prohibitive for the Bayesian method. For the motif-based method, no reconstruction was obtained within the 80-h time limit for any tested size from $N=1000$ to $5000$, so no RMI values could be reported. We further compare with the Bayesian reconstruction method on the nine empirical datasets reported in Table~\ref{tab:empirical_results}, using the pairwise projections of these hypergraphs for both methods. As shown in Table~\ref{tab:bayesian_empirical}, our method achieves higher RMI and shorter runtimes on all four datasets for which the Bayesian method finishes. On the remaining five datasets, the Bayesian runs exceed the 80-GB memory limit and terminate without producing a reconstruction, because the number of maximal cliques grows rapidly on denser empirical projections, even at moderate $N$.

\section{Conclusion}
\label{sec:conclusion}
Here we derived an information theoretic objective for hypergraph reconstruction in which a hyperedge is retained only when the compression it affords exceeds the cost of specifying it, and showed that the locality of this objective supports a seed-centered search that scales to hundreds of thousands of observed edges without a maximum interaction order. The same objective yields a detectability boundary, reducible to the size--density condition of Eq.~\eqref{eq:leading_detectability_boundary} in the sparse regime, which coincides at leading order with the statistical limit for single-interaction recovery in the large-interaction limit and predicts which hyperedges are recovered when many coexist. Our method can successfully recover planted higher-order structure in a wide range of synthetic hypergraphs, and performs well compared to existing methods on hypergraph reconstruction tests in both synthetic and empirical benchmarks, with only a small fraction of the runtime. 

There are a number of ways to extend our approach to future work. While our methodology uses the observed pair density $\rho$ as a global background model, this choice may become conservative in regimes where higher-order projections from the latent hypergraph $H$ contribute a large share of the observed pairwise graph $G$. Appendix~\ref{app:rho_sensitivity} examines this sensitivity and suggests that an optional rescaling guided by the description length objective may improve recovery in some regimes, at the cost of additional reconstruction runs. More generally, heterogeneous or community-structured background models could better account for systematic variation in pairwise density across the network, at the cost of additional computational expense. The observation model also extends directly to weighted or repeated edges, with the Bernoulli KL gain replaced by the corresponding likelihood gain relative to an appropriate background model. Finally, the algorithm locality allows seed-centered searches to be distributed across compute nodes for large-scale parallelization in future implementations.  

\section*{Acknowledgments}
\vspace{-\baselineskip}
A.K. acknowledges support from the HKU-100 Start Up Fund.


\begin{thebibliography}{10}
\bibitem{runtime_note}
All inference runs were performed on a single compute node with 4 CPU cores allocated and 80 GB RAM. Reported runtimes are wall-clock times.
\end{thebibliography}

\clearpage
\appendix
\raggedbottom
\onecolumngrid

\section{Hypergraph-conditioned data code}
\label{app:hypergraph-conditioned-data-code}
Here we derive Eq.~\eqref{eq:main_exact_final_identity} of the main text and the monotonicity property used in Sec.~\ref{sec:objective}. With $Y_{ij}$, $\rho$, $M$, $m$, and the attribution $z$ as defined in Sec.~\ref{sec:structural}, define for each $h\in H$ the owned pair set and owned counts
\begin{equation*}
    \mathcal{P}_{\mathrm{own}}(h)=\{(i,j): i<j,\ z_{ij}=h\},
    \qquad
    T_h = |\mathcal{P}_{\mathrm{own}}(h)|,
    \qquad
    e_h=\sum_{(i,j)\in\mathcal{P}_{\mathrm{own}}(h)}Y_{ij}.
\end{equation*}
Similarly, define the background counts
\begin{equation*}
    T_0=\sum_{i<j}\mathbf{1}\{z_{ij}=0\},
    \qquad
    e_0=\sum_{i<j}\mathbf{1}\{z_{ij}=0\}Y_{ij}.
\end{equation*}
Then 
\begin{equation*}
    m=e_0+\sum_{h\in H}e_h, \qquad M=T_0+\sum_{h\in H}T_h.
\end{equation*}
Under the attribution-based Bernoulli model of the main text, the conditional likelihood factorizes as
\begin{equation*}
    P(G\mid H,z,\rho,\{\theta_h\})=\prod_{i<j}\left[\bigl(\rho^{Y_{ij}}(1-\rho)^{1-Y_{ij}}\bigr)^{\mathbf{1}\{z_{ij}=0\}}\cdot\prod_{h\in H}\bigl(\theta_h^{Y_{ij}}(1-\theta_h)^{1-Y_{ij}}\bigr)^{\mathbf{1}\{z_{ij}=h\}}\right].
\end{equation*}
The data description length is
\begin{equation*}
    L_{\mathrm{data}}(G\mid H,z,\rho,\{\theta_h\})=-\log P(G\mid H,z,\rho,\{\theta_h\}),
\end{equation*}
and here we adopt the standard profile/plug-in convention
\begin{equation}
    \label{eq:profile_def}
    L_{\mathrm{data}}(G\mid H,z,\rho):=\min_{\{\theta_h\in[0,1]\}}L_{\mathrm{data}}(G\mid H,z,\rho,\{\theta_h\}).
\end{equation}
We use the convention $0\log0=0$, so that the minimum in Eq.~\eqref{eq:profile_def} is attained on the closed interval even when $\hat p_h\in\{0,1\}$, as for a fully observed clique ($\hat p_h=1$). The quantities $T_h$, $e_h$, and $\hat p_h=e_h/T_h$ are defined with respect to the current $z$, but for simplicity we suppress this dependence in the notation unless it is needed. 

By exclusive pair attribution, the negative log-likelihood decomposes over the disjoint sets of attributed node pairs induced by $z$, thus
\begin{equation*}
    L_{\mathrm{data}}(G\mid H,z,\rho,\{\theta_h\})=L_0(G\mid z,\rho)+\sum_{h\in H}L_h(G\mid z,\theta_h),
\end{equation*}
where 
\begin{equation*}
    \begin{aligned}
    L_0(G\mid z,\rho)&=-e_0\log\rho-(T_0-e_0)\log(1-\rho),\\
    L_h(G\mid z,\theta_h)&=-e_h\log \theta_h-(T_h-e_h)\log(1-\theta_h).
    \end{aligned}
\end{equation*}
Similarly, for the baseline code we can write
\begin{equation*}
    L_{\mathrm{base}}(G\mid \rho)=L_{0,\mathrm{base}}(G\mid z,\rho)+\sum_{h\in H}L_{h,\mathrm{base}}(G\mid z,\rho),
\end{equation*}
where
\begin{equation*}
    \begin{aligned}
    L_{0,\mathrm{base}}(G\mid z,\rho)&=-e_0\log\rho-(T_0-e_0)\log(1-\rho),\\
    L_{h,\mathrm{base}}(G\mid z,\rho)&=-e_h\log\rho-(T_h-e_h)\log(1-\rho).
    \end{aligned}
\end{equation*}
Since $L_{0,\mathrm{base}}(G\mid z,\rho)=L_0(G\mid z,\rho)$, we then have
\begin{equation}
    \label{eq:rewrite_step}
    L_{\mathrm{data}}(G\mid H,z,\rho,\{\theta_h\})=L_{\mathrm{base}}(G\mid \rho)+\sum_{h\in H}\Bigl[L_h(G\mid z,\theta_h)-L_{h,\mathrm{base}}(G\mid z,\rho)\Bigr], 
\end{equation}
where for each $h\in H$, we have
\begin{equation}
    \label{eq:local_difference}
    L_h(G\mid z,\theta_h)-L_{h,\mathrm{base}}(G\mid z,\rho) = e_h\log\frac{\rho}{\theta_h} + (T_h-e_h)\log\frac{1-\rho}{1-\theta_h}.
\end{equation}
Since $L_{h,\mathrm{base}}(G\mid z,\rho)$ is independent of $\theta_h$, minimizing this difference over $\theta_h$ is equivalent to maximizing the Bernoulli log-likelihood 
\begin{equation*}
    \ell_h(\theta_h)=e_h\log \theta_h+(T_h-e_h)\log(1-\theta_h).
\end{equation*}

Differentiating gives 
\begin{equation*}
    \partial_{\theta_h}\ell_h=\frac{1}{\ln 2}\left[\frac{e_h}{\theta_h}-\frac{T_h-e_h}{1-\theta_h}\right],
    \qquad
    \partial_{\theta_h}^2\ell_h=-\frac{1}{\ln 2}\left[\frac{e_h}{\theta_h^2}+\frac{T_h-e_h}{(1-\theta_h)^2}\right]<0,
\end{equation*}
so for $0<e_h<T_h$ the maximizer is $\theta_h^\star=\hat p_h=e_h/T_h$. If $e_h=0$ ($e_h=T_h$), $\ell_h$ is monotonically decreasing (increasing) in $\theta_h$ and is maximized at $\theta_h=0$ ($\theta_h=1$), which again equals $\hat p_h$. Substituting $\theta_h=\hat p_h$ into Eq.~\eqref{eq:local_difference} yields
\begin{align*}
    \min_{\theta_h}\Bigl[L_h(G\mid z,\theta_h)-L_{h,\mathrm{base}}(G\mid z,\rho)\Bigr]
    &=e_h\log\frac{\rho}{\hat p_h}+(T_h-e_h)\log\frac{1-\rho}{1-\hat p_h}\nonumber\\
    &=-T_h\left[\hat p_h\log\frac{\hat p_h}{\rho}+(1-\hat p_h)\log\frac{1-\hat p_h}{1-\rho}\right]\nonumber\\
    &=-\,T_hD_{\mathrm{KL}}(\hat p_h\|\rho).
\end{align*}
Applying this independently to each $h$ in Eq.~\eqref{eq:rewrite_step} yields the exact identity
\begin{equation}
    \label{eq:exact_final_identity}
    L_{\mathrm{data}}(G\mid H,z,\rho)=L_{\mathrm{base}}(G\mid \rho)-\sum_{h\in H}T_h\,D_{\mathrm{KL}}(\hat p_h\|\rho),
\end{equation}
which is Eq.~\eqref{eq:main_exact_final_identity} of the main text.
 
For fixed $e_h$, define
\begin{equation*}
    g(n)=n\,D_{\mathrm{KL}}\!\left(\frac{e_h}{n}\middle\|\rho\right),
    \qquad
    g'(n)=\log\frac{1-e_h/n}{1-\rho},
\end{equation*}
where $g'(n)$ is non-positive whenever $e_h/n\ge\rho$. Because a pair can be owned by $h$ only if both of its endpoints lie in $h$, the owned pairs of $h$ are a subset of its internal pairs, so $T_h\le P_h=\binom{|h|}{2}$. Since $e_h/n$ decreases in $n$, the condition $\tilde p_h=e_h/P_h\ge\rho$ implies that $e_h/n\ge\rho$ throughout the interval $[T_h,P_h]$, and hence
\begin{equation*}
    P_h\,D_{\mathrm{KL}}(\tilde p_h\|\rho)\le T_h\,D_{\mathrm{KL}}(\hat p_h\|\rho).
\end{equation*}
Replacing $T_hD_{\mathrm{KL}}(\hat p_h\|\rho)$ by $\mathrm{gain}(h;z)=P_hD_{\mathrm{KL}}(\tilde p_h\|\rho)$ in Eq.~\eqref{eq:exact_final_identity} yields an upper bound on the description length conditional on $z$, with equality when retained hyperedges do not overlap ($T_h=P_h$ for all $h$). The bound requires $\tilde p_h\ge\rho$ for every retained $h$, which the search enforces both when accepting candidates (Sec.~\ref{sec:objective}) and during reassignment (Appendix~\ref{sec:local_evaluation}), so this holds for every hypergraph the search visits.

\clearpage
\section{MAP interpretation of the MDL objective}
\label{app:MAP}
The code of Eq.~\eqref{eq:Lmodel_H} contains no separate code for the number or ordering of the retained hyperedges. We show that it arises, up to an $H$-independent normalization constant, from an independent-inclusion prior over subsets of the candidate universe. Let $\mathcal{U}_N=\{h\subseteq V:3\le |h|\le N\}$ be the finite set of valid hyperedges, and associate each $h\in\mathcal{U}_N$ with a weight
\begin{equation*}
    q(h)=2^{-L_{\mathrm{model}}(h)}.
\end{equation*}
The induced distribution over subsets $H\subseteq\mathcal{U}_N$ is
\begin{equation}
    \label{eq:indep_inclusion_prior}
    P_{\mathrm{model}}(H)=\frac{1}{Z_N}\prod_{h\in H}q(h),
    \qquad
    Z_N=\prod_{h\in\mathcal{U}_N}\bigl(1+q(h)\bigr),
\end{equation}
which defines a normalized prior over unordered hyperedge sets of every cardinality. Its normalization constant is bounded and independent of $(H,z)$. To see this, summing $q(h)$ over $\mathcal{U}_N$ and grouping hyperedges by size, using 
\begin{equation*}
    q(h)=2^{-L_{\mathrm{model}}(|h|)}=\binom{N}{s}^{-1}\frac{N}{s(s-1)(N-1)}
\end{equation*}
from Eq.~\eqref{eq:Lmodel_h}, gives
\begin{equation*}
    \sum_{h\in\mathcal{U}_N}q(h)
    =\sum_{s=3}^{N}\binom{N}{s}\frac{N}{s(s-1)(N-1)}\binom{N}{s}^{-1}
    =\frac{N}{N-1}\sum_{s=3}^{N}\frac{1}{s(s-1)}
    =\frac{N-2}{2(N-1)}<\frac12,
\end{equation*}
using the telescoping identity 
\begin{equation*}
    \sum_{s=3}^{N}\frac{1}{s(s-1)}=\frac12-\frac1N.
\end{equation*}
Hence, using $\log(1+x)\le x/\ln2$ for $x\ge0$,
\begin{equation*}
    C_N:=\log Z_N=\sum_{h\in\mathcal{U}_N}\log\bigl(1+q(h)\bigr)\le\frac{1}{\ln 2}\sum_{h\in\mathcal{U}_N}q(h)<\frac{1}{2\ln 2}\approx0.72<1\ \text{bit},
\end{equation*}
uniformly in $N$. The exact structural code length is therefore 
\begin{equation*}
    -\log P_{\mathrm{model}}(H)=L_{\mathrm{model}}(H)+C_N.
\end{equation*}
At fixed $N$, $C_N$ is independent of $(H,z)$ and cancels from the description length differences and the maximum a posteriori (MAP) criterion. Thus, neither the number nor the ordering of the retained hyperedges requires a separate code. The MDL objective can thus be interpreted as a MAP model-selection criterion under the prior of Eq.~\eqref{eq:indep_inclusion_prior}. 

Combining this structural prior with the likelihood $P_{\mathrm{data}}(G\mid H,z,\rho)=2^{-L_{\mathrm{data}}(G\mid H,z,\rho)}$ of Eq.~\eqref{eq:profile_def} gives
\begin{equation*}
\begin{aligned}
    \arg\min_{H,z}\left[L_{\mathrm{model}}(H)+L_{\mathrm{data}}(G\mid H,z,\rho)\right]
    &=\arg\min_{H,z}\left[L_{\mathrm{model}}(H)+L_{\mathrm{data}}(G\mid H,z,\rho)+C_N\right] \\
    &=\arg\min_{H,z}\left[-\log P_{\mathrm{model}}(H)-\log P_{\mathrm{data}}(G\mid H,z,\rho)\right] \\
    &=\arg\max_{H,z}P_{\mathrm{model}}(H)P_{\mathrm{data}}(G\mid H,z,\rho).
\end{aligned}
\end{equation*}
Thus, minimizing the MDL objective is equivalent to selecting the MAP reconstruction under this prior. (Because $L_{\mathrm{data}}$ is profiled over $\{\theta_h\}$, $P_{\mathrm{data}}$ is a maximized rather than a normalized likelihood, and no prior is placed on $z$ (see Sec.~\ref{sec:attribution}). The correspondence is therefore a penalized profile-likelihood criterion rather than an \emph{exact} MAP estimator, although the omitted parameter cost is $O(\log|h|)$ bits per hyperedge (Sec.~\ref{sec:attribution}), which is lower order than $L_{\mathrm{model}}(h)=\Theta(|h|\log(N/|h|))$.)

\clearpage
\section{Detectability limit for reconstruction of single interactions}
\label{app:information_boundary}

In this Appendix we prove the optimality claim of Sec.~\ref{sec:information_boundary}. We consider a graph that contains either no interaction or exactly one interaction of known size $s$, and we ask under what conditions the node set of that interaction can be recovered. The relevant quantity is the Kullback--Leibler divergence between the planted and background models,
\begin{equation*}
    I_s=D_{\rm KL}(\mathbb P_S\Vert\mathbb P_0)=P_s\,D_{\rm KL}(c\Vert\rho),
\end{equation*}
which measures, in bits, how distinguishable a graph containing the planted interaction is from a graph without it (Sec.~\ref{app:C_setup}). We establish three results:
\begin{enumerate}[label=(\roman*)]
    \item No method achieves reliable recovery if $I_s<(1-\epsilon)\log\binom Ns$ for some fixed $\epsilon>0$.
    \item A thresholded likelihood scan achieves reliable recovery if $I_s>(1+\epsilon)\log\binom Ns$.
    \item The MDL criterion of the main text also achieves reliable recovery if $I_s>(1+\epsilon)\log\binom Ns$, without requiring knowledge of the internal density $c$.
\end{enumerate}
These results rest on a counting argument. Specifying one node set of size $s$ among the $\binom Ns$ candidates requires $\log\binom Ns$ bits, so the planted and background models must differ by at least this much information for the interaction to be identified. The MDL code assigns a hyperedge of size $s$ a structural cost of
\begin{equation*}
    L_{\rm model}(s)=[1+o(1)]\log\binom Ns
\end{equation*}
(Eq.~\eqref{eq:model_cost_asymptotic}), so its acceptance threshold coincides with the statistical limit at leading order without any tuning. Sections~\ref{app:C_setup}--\ref{app:C_mdl} state the results and outline the arguments, with the proof of Proposition~2 given in full in Sec.~\ref{app:C_mdl} and Sec.~\ref{app:C_proofs} giving the complete proof of Proposition~1.

\subsection{Setting}
\label{app:C_setup}
Denote the set of candidate node sets of size $s\ge3$ as
\begin{equation*}
    \mathcal U_{N,s}=\{S\subseteq V:|S|=s\},
\end{equation*}
and let $P_s=\binom s2$. We compare two models for the observed graph:
\begin{itemize}
    \item \emph{Background model} $\mathbb P_0$: each node pair is an edge independently with probability $\rho$.
    \item \emph{Planted model} $\mathbb P_S$: each of the $P_s$ pairs inside $S$ is an edge with probability $c>\rho$, and every other pair is an edge with probability $\rho$, all independently.
\end{itemize}
The two models differ only on the pairs inside $S$, and pairs are independent, so the divergence between them is additive over these pairs:
\begin{equation*}
    I_s=D_{\rm KL}(\mathbb P_S\Vert\mathbb P_0)=P_s\,D_{\rm KL}(c\Vert\rho).
\end{equation*}

An estimator returns either $\hat S=\varnothing$, indicating that no interaction is present, or a set $\hat S\in\mathcal U_{N,s}$, and we call such an estimator \emph{reliable} if it satisfies both of the following:
\begin{itemize}
    \item \emph{No false detection:} the probability of reporting an interaction under the background model vanishes.
    \item \emph{Accurate localization:} under the planted model, the expected fraction of misidentified nodes vanishes, uniformly over $S$.
\end{itemize}
Formally, these conditions are given by
\begin{equation}
    \mathbb P_0(\hat S\neq\varnothing)\longrightarrow0,
    \qquad
    \sup_{S\in\mathcal U_{N,s}}
    \frac{\mathbb E_S|\hat S\triangle S|}{s}\longrightarrow0.
    \label{eq:oneint_reliable}
\end{equation}
Here $\mathbb E_S$ denotes expectation under $\mathbb P_S$, $\triangle$ denotes the symmetric difference, and we set $|\varnothing\triangle S|=s$. Throughout, we take $N\to\infty$ with $s\to\infty$ and $s=o(N)$.

\subsection{The information limit}
\label{app:C_limit}
\noindent\textbf{Proposition 1.} Let $\epsilon>0$ be fixed.
\begin{enumerate}[label=(\alph*)]
    \item If $I_s<(1-\epsilon)\log\binom Ns$, no estimator is reliable.
    \item If $I_s>(1+\epsilon)\log\binom Ns$, and $c$, $1-c$, and $c-\rho$ are bounded below by a positive constant, the thresholded likelihood scan defined below is reliable.
\end{enumerate}

\noindent\emph{Outline of (a).}
\begin{itemize}
    \item A reliable estimator must, with high probability, return a set that differs from $S$ in only $o(s)$ nodes.
    \item The number of sets within this distance of any given set is $(N/s)^{o(s)}$, a negligible fraction of the $\binom Ns\approx(N/s)^s$ candidates.
    \item Reliable recovery is therefore essentially equivalent to identifying $S$ among $\binom Ns^{1-o(1)}$ distinct alternatives, and the data processing inequality shows that this requires $I_s\ge[1-o(1)]\log\binom Ns$.
\end{itemize}

\noindent\emph{Outline of (b).} \\
For each candidate $T$, the scan computes the log-likelihood ratio and its maximizer,
\begin{equation*}
    \Lambda_T=\log\frac{\mathbb P_T(G)}{\mathbb P_0(G)},
    \qquad
    \tilde S\in\argmax_{T\in\mathcal U_{N,s}}\Lambda_T,
\end{equation*}
and returns $\tilde S$ if $\Lambda_{\tilde S}$ exceeds the threshold
\begin{equation*}
    t_N=\left(1+\frac\epsilon2\right)\log\binom Ns,
\end{equation*}
and $\varnothing$ otherwise. We can then check the two conditions for reliability:
\begin{itemize}
    \item \emph{False detection.} Under the background model, each candidate exceeds the threshold with probability at most $2^{-t_N}$, so a union bound over the $\binom Ns$ candidates shows that false detections vanish.
    \item \emph{Localization.} Under the planted model, a candidate that differs from $S$ in $k$ nodes has a log-likelihood deficit relative to $S$ of order $(k/s)I_s$, whereas the number of such candidates is approximately
    \begin{equation*}
        \left(\frac Ns\right)^k=2^{(k/s)\log\binom Ns}.
    \end{equation*}
    When $I_s$ exceeds $\log\binom Ns$ by the factor $1+\epsilon$, the deficit outweighs the number of competitors for every $k$ of order $s$, so the selected set is close to $S$.
\end{itemize}

Dividing by $s$, the boundary $I_s=\log\binom Ns$ becomes
\begin{equation*}
    \frac{s-1}{2}\,D_{\rm KL}(c\Vert\rho)=[1+o(1)]\log\frac Ns,
\end{equation*}
which is the large-$s$ form of Eq.~\eqref{eq:scaling_retention_full}. The single-interaction problem is the planted dense subgraph (hidden community) problem~\cite{butucea2013detection,ariascastro2014community,hajek2017information}, and this boundary coincides with the weak-recovery threshold of Ref.~\cite{hajek2017information}. 
Because our reliability criterion also requires the false detection probability to vanish, we give a self-contained proof in Sec.~\ref{app:C_proofs}.

\subsection{Attainment by the MDL criterion}
\label{app:C_mdl}
For a candidate $T$, define its number of internal edges, internal edge density, and gain as
\begin{equation*}
    e(T)=\sum_{\{i,j\}\subseteq T}Y_{ij},
    \qquad
    \tilde p_T=\frac{e(T)}{P_s},
    \qquad
    \operatorname{gain}(T)=P_s\,D_{\rm KL}(\tilde p_T\Vert\rho).
\end{equation*}
These are the single-interaction counterparts of $e_h$, $\tilde p_h$, and $\operatorname{gain}(h;z)$ in the main text. By Eq.~\eqref{eq:additive_total_main}, the description lengths of the two hypotheses are
\begin{equation*}
    L(\varnothing)=L_{\rm base},
    \qquad
    L(\{T\})=L_{\rm base}+L_{\rm model}(s)-\operatorname{gain}(T).
\end{equation*}
Since $L_{\rm model}(s)$ is the same for every candidate, the MDL criterion will select the candidate $T$ with the largest gain, and accept it if and only if $\tilde p_T>\rho$ and $\operatorname{gain}(T)>L_{\rm model}(s)$.\\

\noindent\textbf{Proposition 2.} Under the conditions of Proposition~1(b), the MDL criterion is reliable.

We can prove this by verifying the two conditions of Eq.~\eqref{eq:oneint_reliable}, as before:

\begin{itemize}
    \item \emph{False detection.}
    \begin{itemize}
        \item For a fixed $T$, the acceptance event $\{\tilde p_T>\rho,\ \operatorname{gain}(T)>L_{\rm model}(s)\}$ is equivalent to $\{\tilde p_T>a_s\}$, where $a_s>\rho$ solves
        \begin{equation*}
            P_s\,D_{\rm KL}(a_s\Vert\rho)=L_{\rm model}(s).
        \end{equation*}
        If no such $a_s\le1$ exists, the event is empty.
        \item By the Chernoff bound, $\mathbb P_0(\tilde p_T>a_s)\le2^{-P_sD_{\rm KL}(a_s\Vert\rho)}=2^{-L_{\rm model}(s)}$.
        \item A union bound over all $\binom Ns$ candidates and Eq.~\eqref{eq:Lmodel_h} then give
        \begin{equation*}
            \mathbb P_0(\hat S\neq\varnothing)\le\binom Ns2^{-L_{\rm model}(s)}=\frac{N}{s(s-1)(N-1)}\longrightarrow0.
        \end{equation*}
    \end{itemize}

    \item \emph{Localization.}
    \begin{itemize}
        \item For $\tilde p_T>\rho$, both $\operatorname{gain}(T)$ and $\Lambda_T$ are increasing in $e(T)$. The MDL criterion therefore selects the same set $\tilde S$ as the likelihood scan, and by Proposition~1(b) this set differs from $S$ in $o(s)$ nodes.
        \item It remains to show that $\tilde S$ is accepted. By the law of large numbers, $\operatorname{gain}(S)/I_s\to1$ in probability. Since $\operatorname{gain}(\tilde S)\ge\operatorname{gain}(S)$, the hypothesis $I_s>(1+\epsilon)\log\binom Ns$ of Proposition~1(b) (restated as Eq.~\eqref{eq:achievability_gap} in Sec.~\ref{app:C_proofs}) and Eq.~\eqref{eq:model_cost_asymptotic} give, with probability tending to one,
        \begin{equation*}
            \operatorname{gain}(\tilde S)\ge[1-o(1)]\,I_s>(1+\epsilon)[1-o(1)]\,L_{\rm model}(s)>L_{\rm model}(s).
        \end{equation*}
    \end{itemize}
\end{itemize}

We can now note that the likelihood scan requires the value of $c$ to compute $\Lambda_T$, as well as a threshold chosen in relation to $\log\binom Ns$. However, the MDL criterion does not have these requirements, as its threshold is the structural cost $L_{\rm model}(s)$, which acts as a correction for testing all $\binom Ns$ candidates and falls at the information limit by construction.

\subsection{Proofs of Proposition 1}
\label{app:C_proofs}

\subsubsection*{\underline{\textbf{Part (a)}}}
For $S,T\in\mathcal U_{N,s}$, define the distance
\begin{equation*}
    d(S,T)=|S\setminus T|=\frac{|S\triangle T|}{2},
\end{equation*}
the number of nodes that must be exchanged to turn one set into the other. The number of sets within distance $k$ of a given set is
\begin{equation*}
    B(k)=\sum_{j=0}^{k}\binom sj\binom{N-s}{j}.
\end{equation*}

\noindent\textbf{Step 1.} Suppose an estimator is reliable, and define
\begin{equation*}
    a_N=\sup_{S}\frac{\mathbb E_S|\hat S\triangle S|}{s}\longrightarrow0,
    \qquad
    k_N=\max\bigl\{1,\lceil s\sqrt{a_N}\,\rceil\bigr\}=o(s).
\end{equation*}
Let $\mathcal E_S=\{\hat S\neq\varnothing,\ d(\hat S,S)\le k_N\}$ be the event that the output is nonempty and close to $S$. On the complement $\mathcal E_S^c$, the normalized loss $|\hat S\triangle S|/s$ is large, such that an empty output contributes a loss of exactly one and a nonempty output outside the ball contributes a loss greater than $2k_N/s$. Markov's inequality therefore gives, uniformly in $S$,
\begin{equation*}
    \mathbb P_S(\mathcal E_S^c)\le a_N+\frac{a_Ns}{2k_N}\le a_N+\frac{\sqrt{a_N}}{2}\longrightarrow0.
\end{equation*}

\noindent\textbf{Step 2.} Now, average the probability of $\mathcal E_S$ over candidates under each model,
\begin{equation*}
    \bar p=\binom Ns^{-1}\sum_S\mathbb P_S(\mathcal E_S),
    \qquad
    \bar q=\binom Ns^{-1}\sum_S\mathbb P_0(\mathcal E_S).
\end{equation*}
\begin{itemize}
    \item By Step 1, $\bar p=1-o(1)$.
    \item Any nonempty output lies within distance $k_N$ of exactly $B(k_N)$ candidates, so at most $B(k_N)$ of the events $\mathcal E_S$ hold simultaneously. Hence $\bar q\le B(k_N)/\binom Ns$.
    \item Applying the data processing inequality to the indicator of $\mathcal E_S$ gives $I_s\ge D_{\rm KL}\bigl(\mathbb P_S(\mathcal E_S)\Vert\mathbb P_0(\mathcal E_S)\bigr)$ for each $S$.
    \item Averaging over $S$ using the convexity of the KL divergence, and using $D_{\rm KL}(p\Vert q)\ge p\log(1/q)-1$ for binary distributions, we obtain
    \begin{equation*}
        I_s\ge D_{\rm KL}(\bar p\Vert\bar q)
        \ge \bar p\log\frac1{\bar q}-1
        \ge[1-o(1)]\log\frac{\binom Ns}{B(k_N)}.
    \end{equation*}
\end{itemize}

\noindent\textbf{Step 3.} Using $\binom nj\le(en/j)^j$ and $k_N=o(s)$,
\begin{equation*}
    \log B(k_N)
    \le \log(k_N+1)+k_N\log\!\frac{e^2sN}{k_N^2}
    =k_N\log\frac Ns+O\!\left(k_N\log\frac s{k_N}\right)
    =o\!\left(s\log\frac Ns\right).
\end{equation*}
Since $\log\binom Ns\ge s\log(N/s)$, this is $o(\log\binom Ns)$. Combining with Step 2, every reliable estimator satisfies
\begin{equation*}
    I_s\ge[1-o(1)]\log\binom Ns,
\end{equation*}
which proves part (a).

\subsubsection*{\underline{\textbf{Part (b)}}}
Suppose
\begin{equation}
    I_s>(1+\epsilon)\log\binom Ns,
    \label{eq:achievability_gap}
\end{equation}
and that $c$, $1-c$, and $c-\rho$ are bounded below by a positive constant. Because the planted and background models differ only on pairs inside $T$, the log-likelihood ratio of a candidate $T$ is
\begin{equation*}
    \Lambda_T=P_s\log\frac{1-c}{1-\rho}+e(T)\,\beta,
    \qquad
    \beta=\log\frac{c(1-\rho)}{\rho(1-c)}>0,
\end{equation*}
which is increasing in $e(T)$. The scan selects $\tilde S\in\argmax_T\Lambda_T$ and returns
\begin{equation*}
    \hat S=
    \begin{cases}
        \tilde S, & \Lambda_{\tilde S}>t_N,\\
        \varnothing, & \text{otherwise},
    \end{cases}
    \qquad
    t_N=\left(1+\frac\epsilon2\right)\log\binom Ns.
\end{equation*}

\noindent\textbf{Step 1 (false detection).} Under $\mathbb P_0$, $\mathbb E_0[2^{\Lambda_T}]=1$ for every $T$. Markov's inequality and a union bound over the $\binom Ns$ candidates then give
\begin{equation*}
    \mathbb P_0(\hat S\neq\varnothing)\le\binom Ns2^{-t_N}=\binom Ns^{-\epsilon/2}\longrightarrow0.
\end{equation*}

\noindent\textbf{Step 2.} Under $\mathbb P_S$, $e(S)\sim{\rm Binomial}(P_s,c)$, so
\begin{equation*}
    \mathbb E_S[\Lambda_S]=I_s,
    \qquad
    \operatorname{Var}_S(\Lambda_S)=P_s\,c(1-c)\,\beta^2.
\end{equation*}
\begin{itemize}
    \item In the stated regime the ratio $\beta/D_{\rm KL}(c\Vert\rho)$ is bounded, so $\operatorname{Var}_S(\Lambda_S)/I_s^2=O(1/P_s)\to0$.
    \item Chebyshev's inequality then gives $\Lambda_S/I_s\to1$ in probability.
    \item Since $I_s>(1+\epsilon)\log\binom Ns>t_N$ by Eq.~\eqref{eq:achievability_gap}, it follows that $\mathbb P_S(\Lambda_S>t_N)\to1$.
\end{itemize}

\noindent\textbf{Step 3.} Consider a candidate $T$ at distance $k=d(S,T)$. The two sets share $\binom{s-k}{2}$ internal pairs, and each has
\begin{equation*}
    m_k=P_s-\binom{s-k}{2}=\frac{k(2s-k-1)}{2}\ge\frac ks P_s
\end{equation*}
pairs not shared with the other. Let $X$ and $Z$ denote the numbers of observed edges on the unshared pairs of $S$ and $T$, respectively. Under $\mathbb P_S$,
\begin{equation*}
    X\sim{\rm Binomial}(m_k,c),
    \qquad
    Z\sim{\rm Binomial}(m_k,\rho),
    \qquad
    \Lambda_T-\Lambda_S=\beta\,(Z-X),
\end{equation*}
so $\Lambda_T\ge\Lambda_S$ requires $Z\ge X$. Fix the intermediate density $\tau_N=c-s^{-1/4}$, which lies between $\rho$ and $c$ for large $s$. The event $Z\ge X$ then requires either $X\le m_k\tau_N$ or $Z\ge m_k\tau_N$. For any fixed $\zeta\in(0,1)$, Chernoff bounds and a union bound over all candidates with $k\ge\zeta s$ give
\begin{equation*}
    \mathbb P_S\bigl(d(\tilde S,S)\ge\zeta s\bigr)
    \le \underbrace{\sum_{k\ge\zeta s}\binom sk
    2^{-m_kD_{\rm KL}(\tau_N\Vert c)}}_{\text{(i)}}
    +\underbrace{\sum_{k\ge\zeta s}\binom sk\binom{N-s}{k}
    2^{-m_kD_{\rm KL}(\tau_N\Vert\rho)}}_{\text{(ii)}}.
\end{equation*}
We bound the two sums separately.
\begin{itemize}
    \item \emph{Sum (i).} Here $D_{\rm KL}(\tau_N\Vert c)=\Theta(s^{-1/2})$ and $m_k=\Theta(s^2)$, so each exponent is of order $s^{3/2}$, which dominates the at most $2^s$ terms. Sum (i) therefore vanishes.
    \item \emph{Sum (ii).} Here $D_{\rm KL}(\tau_N\Vert\rho)/D_{\rm KL}(c\Vert\rho)\to1$, so by $m_k\ge(k/s)P_s$ and Eq.~\eqref{eq:achievability_gap}, for large $N$ the exponent satisfies
    \begin{equation*}
        m_kD_{\rm KL}(\tau_N\Vert\rho)\ge\left(1+\frac\epsilon2\right)\frac ks\log\binom Ns\ge\left(1+\frac\epsilon2\right)k\log\frac Ns.
    \end{equation*}
    The number of candidates at distance $k$ satisfies
    \begin{equation*}
        \log\left[\binom sk\binom{N-s}{k}\right]\le k\log\frac Ns+O(k),
    \end{equation*}
    so each term of (ii) is at most
    \begin{equation*}
        2^{-(\epsilon/2)k\log(N/s)+O(k)}.
    \end{equation*}
    Since $\log(N/s)\to\infty$, the sum over $k\ge\zeta s$ vanishes.
\end{itemize}
Hence $d(\tilde S,S)/s\to0$ in probability, and since this ratio is bounded by one, also in expectation.\\

\noindent \textbf{Conclusion.}
\begin{itemize}
    \item By Step 2 and $\Lambda_{\tilde S}\ge\Lambda_S$, the scan returns $\hat S=\tilde S$ with probability tending to one.
    \item Together with Steps 1 and 3, the scan satisfies both conditions of Eq.~\eqref{eq:oneint_reliable}, which proves part (b).
    \item Parts (a) and (b) together show that the sharp threshold for reliable recovery is $I_s=[1+o(1)]\log\binom Ns$, independently of the MDL objective.
\end{itemize}

\clearpage
\section{Search algorithm}
\label{app:search_algorithm}
 
\subsection{Seed selection}
Pair seeds, used for merge proposals, are drawn from the observed edges currently owned by the pairwise background and ranked using two local neighborhood statistics of the endpoints $u$ and $v$. We use the common neighbor count $C_{uv}$ and the normalized neighborhood overlap $R_{uv}$, defined as
\begin{equation*}
    C_{uv}=|\mathcal{N}(u)\cap\mathcal{N}(v)|,
    \qquad
    R_{uv}=\frac{C_{uv}}{|(\mathcal{N}(u)\cup\mathcal{N}(v))\setminus\{u,v\}|},
\end{equation*}
where $\mathcal{N}(\cdot)$ denotes the neighborhood in $G$ and $R_{uv}=0$ when the denominator vanishes. Candidate pairs are ranked separately in descending order by $C_{uv}$ and $R_{uv}$ and then ordered by the combined rank score
\begin{equation*}
    \mathrm{score}(u,v)=\mathrm{rank}(C_{uv})+0.5\,\mathrm{rank}(R_{uv}),
\end{equation*}
with smaller scores receiving higher priority.
Hyperedge seeds, used for both merge and split proposals, are drawn from the currently retained hyperedges and prioritized by recent activity, defined as having been created or modified within the last \texttt{recent\_window} outer iterations (Table~\ref{tab:default_params}).
For pair seeds, computing $(C_{uv},R_{uv})$ over all observed pairs at every iteration would be costly on large graphs, so 
we instead evaluate these ranking statistics on a pool of observed pairs sampled uniformly at random, of size at most a fixed multiple of the pair seed budget (five times the budget by default). The ranked portion of the budget is drawn from this pool, and the random portion from the remaining observed pairs, so that random exploration is not confined to the pool.
In both the pair and hyperedge seed pools, a fixed fraction of the seed budget ($20\%$ by default, Table~\ref{tab:default_params}) is reserved for uniformly sampled seeds rather than prioritized ones. When the merge or split stage of an outer iteration fails to improve $L$ beyond the tolerance of Appendix~\ref{app:termination}, this random fraction is multiplied by the stall multiplier in that stage until it improves again, while the total seed budget remains fixed. This shifts search effort toward exploration once the prioritized regions stop yielding improvements.
 
\subsection{Merge proposals}
Each selected merge seed generates candidate hyperedges in three ways: (a) by adding a single node adjacent to the seed (every such node is tried), (b) by adding several nodes through a beam search, or (c) by merging with a nearby retained hyperedge (hyperedge seeds only). In each case the candidate replaces the seed.

The beam search starts from the seed node set $B_0$ and grows it by one node at a time, keeping at each step only a fixed number of the most promising sets to grow further, so that sets differing from the seed by several nodes can be found without exploring all of them. Concretely, for every set $B$ currently kept, the nodes outside $B$ are ranked by their number of neighbors within $B$, $d_B(v)=|\mathcal{N}(v)\cap B|$, and the top $m_{\rm add}$ extensions $B\cup\{v\}$ are generated. The set $B$ itself also remains a candidate, so that a set can stop growing. All sets generated at this step are ranked in descending order by $\mathrm{gain}(B)-L_{\rm model}(B)$ (the negative of the contribution of $B$ to Eq.~\eqref{eq:additive_optimization}), which is computed as if $B$ owned all of its observed internal edges regardless of their current ownership, and the best $b_{\rm add}$ are kept. Only the newly extended sets are grown at the next step, and a set is not extended once it exceeds the seed by $a_{\max}$ nodes. When no set can be extended further, the top $r_{\rm add}$ sets generated at any step that satisfy $\tilde p_B\ge\rho$ (Sec.~\ref{sec:objective}) are evaluated using the exact expression for $\Delta L$.

For merging, a retained hyperedge $A$ is paired with each retained hyperedge that overlaps $A$ or contains one of the $\texttt{bound\_top}$ outside nodes with the most connections to $A$. Pairs in which one hyperedge contains the other are discarded, and the rest are screened by $\tilde p\ge\rho$ for the union and scored by the exact $\Delta L$. 
 
\subsection{Split proposals}
Each selected hyperedge seed generates split candidates in four ways: (a) by removing a single node, (b) by removing several nodes through a beam search, (c) by deleting the hyperedge and returning its edges to the pairwise baseline, or (d) by cutting it into two hyperedges.

The beam search mirrors the one used for merging but removes nodes instead of adding them. For every set currently kept, up to $m_{\rm rem}$ nodes are proposed for removal, drawn equally from those with the fewest neighbors inside the set and those with the highest ratio of neighbors outside the set to neighbors inside it. The best $b_{\rm rem}$ sets are kept at each step, and a set is not reduced further once it has lost $d_{\max}$ nodes. When no set can be reduced further, the top $r_{\rm rem}$ reduced sets that satisfy $\tilde p\ge\rho$ are evaluated using the exact $\Delta L$. For the cuts in (d), bipartition proposals are generated either by thresholding the Fiedler vector of the induced subgraph at its median or by removing all bridges and separating the largest resulting component from the remaining nodes. Bipartitions are proposed only for hyperedges with at least four nodes, and each part must contain at least two nodes, while a part with exactly two nodes is returned to the pairwise background. Like the other moves, a split is applied only if it lowers $L$.

\subsection{Move acceptance and reassignment}\label{sec:local_evaluation}
Because Eq.~\eqref{eq:additive_optimization} is a sum of per-hyperedge terms, each depending only on the node set of $h$ and on the number $e_h$ of observed edges it owns, the change $\Delta L$ produced by a move involves only the terms of hyperedges that the move creates or removes, together with those whose owned edge count changes. Each move is therefore scored by evaluating these few terms rather than the full objective. Before a proposed move is applied, its $\Delta L$ is recomputed in the current state, since earlier moves in the same batch may have changed this value. Proposed moves in a batch are processed in order of increasing $\Delta L$. Two moves in a batch conflict if they modify the same hyperedge or reassign the same observed edge, and only the first of a conflicting pair in this order is applied. Once a batch of merges or splits has been applied, the ownership of observed edges is updated. An observed edge $(u,v)$ can be owned by the pairwise baseline or by any retained hyperedge containing both $u$ and $v$, and it is moved to the owner under which $L$ is smallest, unless it already has that owner. Moves that would leave a hyperedge with $\tilde p_h<\rho$ are skipped, and a hyperedge left with no owned edges is removed. This update is done first for the edges of the hyperedges modified in the batch, then for the edges of any other hyperedge that lost or gained an edge.

\subsection{Termination}\label{app:termination}
The merge stage (merge followed by reassignment) and the split stage (split followed by reassignment) of an outer iteration are each considered non-improving if they lower $L$ by less than
\begin{equation*}
    \max\bigl(\varepsilon_{\mathrm{abs}},\,\varepsilon_{\mathrm{rel}}|L|\bigr),
\end{equation*}
with $L$ evaluated at the start of the stage. The search stops once both stages have been non-improving for two consecutive outer iterations (the patience in Table~\ref{tab:default_params}), or earlier if an outer iteration applies no move and reassigns no edge.
 
\subsection{Default search parameters}
\label{app:defaults}
Table~\ref{tab:default_params} lists the default parameters used by the search algorithm, matching the defaults of \texttt{Runner} in \texttt{klmdl.py} (\url{https://github.com/zangyingbang/KLMDL-Hypergraph-Inference}). All results in this paper use these values, with the exception of the background scale $\alpha$ in Appendix~\ref{app:rho_sensitivity}.
\begin{table}[h]
    \centering
    \caption{Default search parameters, as set in \texttt{klmdl.py}. (All reported results use these values, apart from the background scale $\alpha$ varied in Appendix~\ref{app:rho_sensitivity}.)}
    \label{tab:default_params}
    \begin{tabular}{llr}
        \hline\hline
        Group & Parameter (code name) & Default \\
        \hline
        Seed budgets & pairwise merge-seed budget per iteration (\texttt{pair\_seeds}) & 100 \\
                     & hyperedge merge-seed budget per iteration (\texttt{nonpair\_seeds}) & 100 \\
                     & hyperedge split-seed budget per iteration (\texttt{split\_seeds}) & 50 \\
                     & top-ranked fraction of each seed budget (\texttt{*\_top\_frac}) & 0.8 \\
                     & stall multiplier on random fraction (\texttt{*\_rand\_mult\_on\_stall}) & 3 \\
                     & pair-seed presampling multiplier (\texttt{pair\_presample\_mult}) & 5 \\
                     & recent-activity window (\texttt{recent\_window}) & 3 \\
        \hline
        Merge proposals
                    & maximum boundary nodes used for hyperedge-merge partner search (\texttt{bound\_top}) & 10 \\
                    & beam-search expansion branching $m_{\rm add}$ (\texttt{bundle\_top\_m}) & 3 \\
                    & beam-search expansion beam width $b_{\rm add}$ (\texttt{bundle\_beam}) & 3 \\
                    & maximum nodes added $a_{\max}$ (\texttt{bundle\_max\_add}) & 10 \\
                    & beam-search expansion candidates returned $r_{\rm add}$ (\texttt{bundle\_return\_top}) & 20 \\
        \hline
        Split proposals
                    & beam-search pruning branching $m_{\rm rem}$ (\texttt{split\_bundle\_top\_m}) & 3 \\
                    & beam-search pruning beam width $b_{\rm rem}$ (\texttt{split\_bundle\_beam}) & 3 \\
                    & maximum nodes removed $d_{\max}$ (\texttt{split\_bundle\_max\_remove}) & 10 \\
                    & beam-search pruning candidates returned $r_{\rm rem}$ (\texttt{split\_bundle\_return\_top}) & 20 \\
                    & bipartition methods (\texttt{split\_methods}) & spectral, bridges \\
        \hline
        Acceptance caps & merge moves applied per batch (\texttt{k\_merge}) & 100 \\
                        & split moves applied per batch (\texttt{k\_split}) & 50 \\
                        & reassignment passes per stage (\texttt{reassign\_passes}) & 1 \\
        \hline
        Stopping & absolute tolerance $\varepsilon_{\rm abs}$ (\texttt{eps\_gain\_abs}) & 10 bits \\
                 & relative tolerance $\varepsilon_{\rm rel}$ (\texttt{eps\_gain\_rel}) & $10^{-5}$ \\
                 & patience (\texttt{patience}) & 2 \\
        \hline\hline
    \end{tabular}
\end{table}

\clearpage
\section{Empirical datasets}
\label{app:datasets}
Six of the nine hypergraphs in Table~\ref{tab:empirical_results} (Languages, Foursquare, Board of Directors, Terrorism, Pollination, and American Revolution) are constructed from public bipartite networks in the Netzschleuder repository~\cite{peixoto2020netzschleuder}, using the convention of Ref.~\cite{young2021reconstruct}. (Under this convention, the larger mode forms the hypergraph vertex set and neighborhoods in the opposite mode define the hyperedges.) Repeated incidences and singleton hyperedges are removed, duplicate hyperedges are collapsed, and the vertex set is restricted to the retained hyperedges. The remaining three coauthorship hypergraphs are constructed from DBLP XML records for publications in 1980, 1985, and 1990~\cite{dblp}, with authors as nodes and each multi-author record represented by its author set, followed by the same cleaning steps. Two-author records form dyads, which are treated as part of the pairwise background exactly as in Sec.~\ref{sec:synthetic}. For every dataset, the observed graph $G$ is generated by retaining each internal pair of each hyperedge independently with probability $p_h\sim U(0.5,1)$, drawn once per hyperedge, and taking the union of retained pairs, so that overlapping contributions to the pairwise projection are merged into a simple graph.

\section{Evaluation of reconstruction accuracy}
\label{app:rmi}
We quantify agreement between the inferred and ground truth hypergraphs using $\mathrm{RMI}$, an asymmetric normalized reduced mutual information, evaluated on the observed edge set $E$ through its induced edge partition. We map each hypergraph to an edge labeling of $E$ by assigning each observed edge a single label. Edge $(u,v)$ receives label $0$ when no retained hyperedge of size at least three contains both endpoints $u$ and $v$, and otherwise receives a positive label $i$ denoting the selected hyperedge $h_i$ containing $\{u,v\}$. When an observed edge is contained in multiple such hyperedges, we select the one with the largest cardinality, breaking ties by hyperedge index. The same rule is applied to the inferred and ground truth hypergraphs so that both labelings are constructed identically, without using the inferred ownership $z$. This yields two labelings $\hat\ell$ (inferred) and $\ell^\star$ (ground truth) on the same sample set $E$, and we report 
\begin{equation*}
    \mathrm{RMI}(\hat\ell;\ell^\star)=\frac{I_{\mathrm{DM}}(\hat\ell;\ell^\star)}{I_{\mathrm{DM}}(\ell^\star;\ell^\star)},
\end{equation*}
where $I_{\mathrm{DM}}$ is the Dirichlet-multinomial (DM) reduced mutual information defined in Refs.~\cite{jerdee2024mutual,jerdee2025RMI}. Compared with standard mutual information, the RMI used here reduces the bias toward overly fine inferred partitions and avoids the additional dependence on the inferred labeling introduced by symmetric normalization. By definition, $\mathrm{RMI}(\hat\ell;\ell^\star)=1$ when the induced partitions coincide up to a permutation of the positive labels, while values near $0$ indicate little shared information and negative values indicate less shared information than expected between unrelated labelings. Because the labeling is defined on $E$, a planted interaction with no observed internal pairs does not enter the evaluation, as such interactions leave no trace in $G$ and cannot be recovered by any method. The interaction-level analysis of Fig.~\ref{fig:fig3}, which is based on Jaccard overlap between planted and inferred node sets, complements this edge-level measure.

\clearpage
\section{Background density sensitivity and rescaling}
\label{app:rho_sensitivity}
\begin{figure}[!tbp]
    \centering
    \includegraphics[width=15cm]{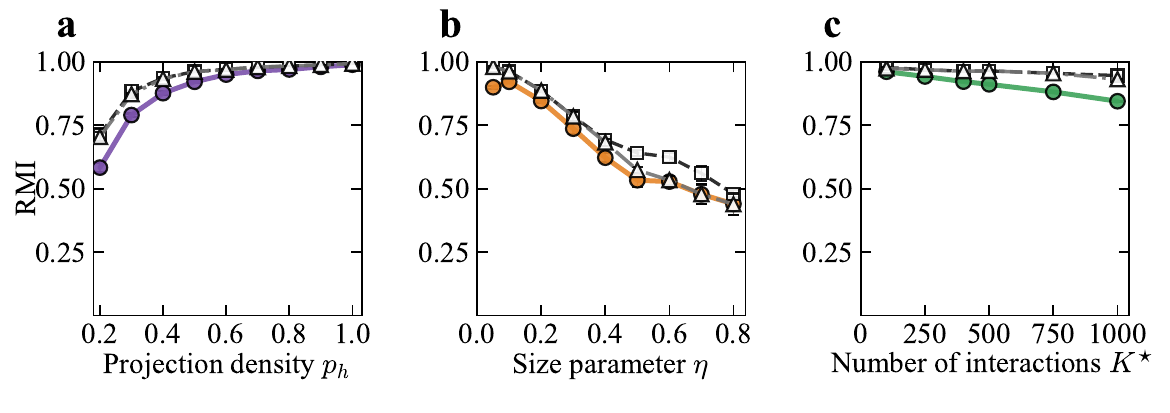}
    \caption{\textbf{Sensitivity to the background density across synthetic simulations.} Reconstruction accuracy for the $N=2000$ experiments of Fig.~\ref{fig:fig2}(a)--(c), plotted against (a) the projection density $p_h$, (b) the size parameter $\eta$, and (c) the number of planted interactions $K^\star$. Filled circles show the default $\alpha=1$ ($\rho_{\rm obj}=\rho_{\rm obs}$), squares the accuracy at the RMI-optimal tested value $\alpha_{\rm best}$, and triangles the accuracy at the value $\alpha_L=\arg\min_\alpha L$ selected by minimizing the description length, where $\alpha_{\rm best}$ and $\alpha_L$ are chosen at each point from the RMI and description length averaged over five independent runs. Markers and error bars show the mean and one standard error over these runs.
    }
    \label{fig:fig5}
\end{figure}

We denote with $\rho_{\rm obj}$ the background density entering the objective, which equals the observed pair density $\rho_{\rm obs}=|E|/\binom{N}{2}$ in the main text. As noted in Sec.~\ref{sec:conclusion}, when a large fraction of observed pairs are projections of latent interactions, $\rho_{\rm obs}$ overestimates the density of the true background, which lowers the gain of every candidate hyperedge and can cause weakly supported interactions to be reduced in size or dropped altogether. Because a Bernoulli code with any background density is a valid code for the observed graph, and because the derivation of Appendix~\ref{app:hypergraph-conditioned-data-code} never uses $\rho=m/M$, so that Eqs.~\eqref{eq:main_exact_final_identity}--\eqref{eq:additive_total_main} hold for any $\rho_{\rm obj}$, the description length remains comparable across different $\rho_{\rm obj}$ (up to the $\log 9\approx3.2$ bits needed to transmit $\alpha$ from the nine value grid below), which is what allows $\alpha$ to be selected by minimizing $L$ below. To assess this background density sensitivity, we scale the objective background density as 
\begin{equation*}
    \rho_{\rm obj}=\alpha\rho_{\rm obs},
    \qquad
    \alpha\in\{1,0.75,0.5,0.25,0.1,0.05,0.01,0.005,0.001\},
\end{equation*}
with $L_{\rm base}$ likewise evaluated at $\rho_{\rm obj}$ so that $L$ is comparable across values of $\alpha$. Table~\ref{tab:rho_sensitivity_load} summarizes the results of the sweep over the number of planted interactions $K^\star$ (the $N=2000$ inset of Fig.~\ref{fig:fig2}(c), where $K^\star/N\in[0.05,0.5]$) at each tested value of $\alpha$. Reducing the objective background density improves RMI throughout the sweep, with the gain over the default increasing from 0.017 at $K^\star=100$ to 0.101 at $K^\star=1000$, consistent with the interpretation that the background at the observed density becomes more conservative at large $K^\star/N$. The RMI-optimal scale is small but finite ($\alpha_{\rm best}\in[0.005,0.05]$), and is never equal to the smallest value tested. The scale selected from the description length also yields essentially the same reconstruction performance as the RMI-optimal scale, with an RMI within 0.005 of the best value in five of six settings and within 0.015 in all settings (Table~\ref{tab:rho_sensitivity_load}, Fig.~\ref{fig:fig5}(c)).
 
\begin{table}[h]
    \caption{\textbf{Sensitivity to the background density.} Results for $N=2000$, $\eta=0.1$, and $p_h\sim U(0.2,1.0)$, with the number of planted interactions $K^\star$ as listed and the background density scaled as $\rho_{\rm obj}=\alpha\rho_{\rm obs}$. $\mathrm{RMI}_{\alpha=1}$ is the default reconstruction accuracy, $\mathrm{RMI}_{\alpha_{\rm best}}$ is the best accuracy over the tested values of $\alpha$ (with the improvement over the default in parentheses), and $\alpha_{\rm best}$ is the value at which the best accuracy is attained. The last column gives $\delta_{\rm sel}=\mathrm{RMI}_{\alpha_{\rm best}}-\mathrm{RMI}_{\alpha_L}$, where $\alpha_L=\arg\min_\alpha L$ is the value selected by minimizing the description length. Dashes indicate $\alpha_L=\alpha_{\rm best}$, and all values are averages over five independent runs.}
    \label{tab:rho_sensitivity_load}
        \begin{ruledtabular}
            \begin{tabular}{ccccc}
            $K^\star$ &
            $\mathrm{RMI}_{\alpha=1}$ &
            $\mathrm{RMI}_{\alpha_{\mathrm{best}}}$ &
            $\alpha_{\mathrm{best}}$ &
            $\delta_{\mathrm{sel}}$ \\
            \hline
            100 & 0.961 & 0.978 (+0.017) & 0.05 & 0.004 ($\alpha_L=0.01$) \\
            250 & 0.942 & 0.969 (+0.027) & 0.005 & -- \\
            400 & 0.922 & 0.963 (+0.041) & 0.005 & $<0.0001$ ($\alpha_L=0.01$) \\
            500 & 0.911 & 0.965 (+0.053) & 0.01 & -- \\
            750 & 0.882 & 0.955 (+0.073) & 0.01 & -- \\
            1000 & 0.845 & 0.946 (+0.101) & 0.005 & 0.014 ($\alpha_L=0.05$) \\
            \end{tabular}
        \end{ruledtabular}
\end{table}
 
Lowering $\rho_{\rm obj}$ has a similar effect in the sweeps over $p_h$ and $\eta$ (Fig.~\ref{fig:fig5}(a,b)), recovering some interactions that are only weakly supported, but interactions whose projected edges are mostly unobserved still cannot be recovered. The final reconstruction accuracy thus does not approach unity for small $p_h$ or large $\eta$. The improvement is small at large $p_h$ and small $\eta$, where the default already recovers nearly all interactions. In all three experiments, the RMI at $\alpha_L$ is close to the RMI at $\alpha_{\rm best}$. These results indicate that searching over $\alpha$ and selecting the tested value that minimizes $L$ is an optional refinement, with the cost of one run of the algorithm per tested value of $\alpha$. On the empirical datasets (Table~\ref{tab:rho_sensitivity_empirical}), it improves the RMI for six of the nine datasets and leaves the three DBLP results essentially unchanged. (For the Languages and Pollination datasets the selected value of $\alpha$ is the smallest one tested ($\alpha_L=0.001$), so extending the tested range could improve these further.) Selecting $\alpha$ in this way uses only the description length, so the gains in Table~\ref{tab:rho_sensitivity_empirical} are available in applications where no ground truth exists.
 
\begin{table}[H]
\caption{\textbf{Sensitivity to the background density on empirical data.} Reconstruction accuracy for the datasets of Table~\ref{tab:empirical_results}, at the default $\alpha=1$ and at the value $\alpha_L$ selected by minimizing the description length over the tested values. Values of $\alpha$ are evaluated in decreasing order over the grid above, with the search stopped after two consecutive values fail to yield a lower description length than the best value found so far. The last column gives the change in accuracy, $\Delta\mathrm{RMI}=\mathrm{RMI}_{\alpha_L}-\mathrm{RMI}_{\alpha=1}$.}
\label{tab:rho_sensitivity_empirical}
    \begin{ruledtabular}
        \begin{tabular}{lcccc}
        Dataset &
        $\mathrm{RMI}_{\alpha=1}$ &
        $\alpha_L$ &
        $\mathrm{RMI}_{\alpha_L}$ &
        $\Delta\mathrm{RMI}$ \\
        \hline
        Languages & 0.861 & 0.001 & 0.935 & +0.074 \\
        Foursquare & 0.865 & 0.05 & 0.900 & +0.035 \\
        Board of Directors & 0.785 & 0.1 & 0.875 & +0.090 \\
        Terrorism & 0.751 & 0.05 & 0.941 & +0.189 \\
        Pollination & 0.923 & 0.001 & 0.989 & +0.066 \\
        American Revolution & 0.921 & 0.005 & 1.000 & +0.079 \\
        DBLP 1980 & 0.976 & 0.005 & 0.971 & -0.005 \\
        DBLP 1985 & 0.968 & 0.01 & 0.964 & -0.003 \\
        DBLP 1990 & 0.969 & 0.05 & 0.970 & +0.001 \\
        \end{tabular}
    \end{ruledtabular}
\end{table}
 
\end{document}